\documentclass[a4paper,11pt]{article}
\usepackage{jheppub}

\usepackage{amsmath, amssymb, mathrsfs}
\usepackage{mathtools}
\usepackage[english]{babel}
\usepackage{graphicx,color,psfrag}
\usepackage{placeins}
\usepackage{subcaption}

\newcommand{\mn}{\ensuremath{{\mu\nu}}}
\newcommand{\dd}{\ensuremath{\mathrm{d}}}
\newcommand{\derp}[2]{\ensuremath{\frac{\partial #1}{\partial #2}}}

\definecolor{vertherbe}{RGB}{58, 157, 35}
\newcommand{\modif}[1]{\textcolor{black}{#1}}

\title{Chiral charge dynamics in Abelian gauge theories at finite temperature}

\newcommand{\addressEPFL}{Institute of Physics, Laboratory of Particle Physics and Cosmology (LPPC), \'Ecole Polytechnique F\'ed\'erale de Lausanne (EPFL), CH-1015 Lausanne, Switzerland.}

\author{Daniel G. Figueroa,}
\emailAdd{daniel.figueroa@epfl.ch}
\author{Adrien Florio}
\emailAdd{adrien.florio@epfl.ch}
\author{and Mikhail Shaposhnikov}
\emailAdd{mikhail.shaposhnikov@epfl.ch}
\affiliation{\addressEPFL}

\date{\today}

\abstract{We study fermion number non-conservation (or chirality breaking) in Abelian gauge theories at finite temperature. We consider the presence of a chemical potential $\mu$ for the fermionic charge, and monitor its evolution with real-time classical lattice simulations. This method accounts for short-scale fluctuations not included in the usual effective magneto-hydrodynamics (MHD) treatment.  We observe a self-similar decay of the chemical potential, accompanied by an inverse cascade process in the gauge field that leads to a production of long-range helical magnetic fields.  We also study the chiral charge dynamics in the presence of an external magnetic field $B$, and extract its decay rate $\Gamma_5 \equiv -{d\log \mu\over dt}$. We provide in this way a new determination of the gauge coupling and magnetic field dependence of the chiral rate, which exhibits a best fit scaling as $\Gamma_5 \propto e^{11/2}B^2$. We confirm numerically the fluctuation-dissipation relation between $\Gamma_5$ and $\Gamma_{\rm diff}$, the Chern-Simons diffusion rate, which was obtained in a previous study. Remarkably, even though we are outside the MHD range of validity, the dynamics observed are in qualitative agreement with MHD predictions. The magnitude of the chiral/diffusion rate is however a factor $\sim 10$ times larger than expected in MHD, signaling that we are in reality exploring a different regime accounting for short scale fluctuations. This discrepancy calls for a revision of the implications of fermion number and chirality non-conservation in finite temperature Abelian gauge theories, though no definite conclusion can be made at this point until hard-thermal-loops are included in the lattice simulations.}


\begin{document}

\maketitle
  \section{Introduction}

  Anomalous processes can be relevant in a large number of phenomena, from high energy particle physics to condensed matter. One of the most well-known applications is the explanation of the $\pi^0 \rightarrow 2\gamma$ decay in quantum electrodynamics (QED), whereas in quantum chromodynamics (QCD) they play a decisive role in the resolution of the $U_A(1)$ problem \cite{tHooft:1976snw,tHooft:1976rip}. The rate at which anomalous processes occur is actually a relevant quantity whenever we are dealing with out-of-equilibrium processes. In highly energetic and dense matter environments like in the early universe, the fluctuations of gauge and scalar fields -- sphalerons \cite{Klinkhamer:1984di}  -- lead to rapid fermion number non-conservation in the Standard Model
   (SM)~\cite{Kuzmin:1985mm},
   and to chirality non-conservation in QCD~\cite{McLerran:1990de}. The $SU(2)$ sphaleron rate is a crucial quantity to assess the viability of electroweak
   baryogenesis~\cite{Kuzmin:1985mm}, and has been extensively studied  across decades, both from a purely analytical side and with the help of numerical simulations. Studies have been carried in the pure  $SU(2)$ theory in Refs.~\cite{Philipsen:1995sg,Ambjorn:1995xm,Arnold:1995bh,Arnold:1996dy,Arnold:1997yb,Moore:1997sn,Bodeker:1998hm,Moore:1998zk,Moore:1999fs,Bodeker:1999gx,Arnold:1999uy} and in the electroweak theory
   e.g.~\cite{Tang:1996qx,Ambjorn:1997jz,Moore:2000mx,DOnofrio:2012phz}, see \cite{DOnofrio:2015gop} for the latest up-to-date prediction using the measured Higgs-mass.

  Anomalous $U(1)$ processes have received some attention, especially in the cosmological context. In the electroweak theory of the SM, the anomaly in the fermionic and/or chiral current actually contains  a $U(1)$
   contribution, which  is associated with the hypercharge field in the Higgs unbroken phase and to the photon field of QED in the Higgs broken phase. As in Abelian gauge theories there are no large gauge transformations, nor vacuum configurations with different Chern-Simons numbers, there is no {\em irreversible} fermion (or chiral) number non-conservation, contrary to the case of non-Abelian theories. This does not prevent however the fermion/chiral number in Abelian theories to be transferred into gauge configurations carrying Chern-Simons number, and to re-appear back again due to changes in the gauge field background. These processes may have an important impact on the problems of baryogenesis~\cite{Giovannini:1997eg,Kamada:2016eeb,Kamada:2016cnb,Fujita:2016igl,Kamada:2018tcs}, magnetic field generation in the early universe~\cite{Joyce:1997uy}, and chiral asymmetry evolution at $\sim \mathrm{MeV} $ temperatures~\cite{Boyarsky:2011uy}.  Anomalous $U(1)$ processes have also received a renewed interest in the quark-gluon plasma community,
  where the chiral magnetic effect~\cite{Vilenkin:1980fu,Fukushima:2008xe} and its potential experimental signatures are being studied, see \cite{Kharzeev:2015znc} for a review.

The above $U(1)$ case have been studied mostly within the framework of magneto-hydrodynamics (MHD), which is an effective description accounting for distance scales exceeding the mean free path of the charged particles involved in the problem, see e.g.~\cite{Brandenburg:2017rcb,Rogachevskii:2017uyc}, or~\cite{Schober:2017cdw} for recent numerical simulations. Despite the relevance of these processes, a full study beyond MHD, taking into account small scale fluctuations in detail, remains to be done. Some attempts in this direction were made in
\cite{Buividovich:2015jfa,Buividovich:2016ulp}, where out-of-equilibrium techniques were implemented. The main limitations of these studies were the intrinsic numerical cost associated with a full-fledged treatment of fermions.  Another approach was initiated in \cite{Figueroa:2017hun}, of which this paper is a natural continuation. To explain the aim of the present research, let us set up a working model and fix notation. We are interested in the study of physics described by scalar electrodynamics coupled to a massless vector-like fermion field $\Psi$, so that our starting lagrangian is
\begin{equation}
	 \mathcal{L} = -{1\over4} F_{\mu\nu}F^{\mu\nu} - {\bar\Psi}\gamma^\mu D_\mu\Psi - (D_\mu\phi)^*(D^\mu\phi) - V(\phi)~,
	 \label{cl}
\end{equation}
where $F_\mn$ is the field strength tensor of the $U(1)$ gauge field $A_\mu$, $D_\mu=\partial_\mu - i e A_\mu$ and
\begin{equation}
  V(\phi) = m^2|\phi|^2 + \lambda|\phi|^4\, .
  \label{eq:potential}
\end{equation}
Taking a positive squared mass $m^2 > 0$, we can chose its value so that the lagrangian~(\ref{cl}) becomes a toy-model for the SM hypercharge sector close to the electroweak phase transition. The chiral fermionic current $J^\mu={\bar\Psi}\gamma^\mu\gamma_5\Psi$ is not conserved at the quantum level, and satisfies the anomaly equation
\begin{align}
	\partial_\mu J_5^\mu &=\frac{e^2}{8\pi^2} F_\mn \tilde{F}^\mn = N_f\partial_\mu K^\mu\,,
	\label{eq:an_rel}
\end{align}
where $\tilde{F}_{\mn}=\frac{1}{2}\epsilon_{\mu\nu\rho\sigma}F^{\rho\sigma}$ is the hodge-dual of $F_\mn$, $N_f$ is the number of flavours and $K^\mu=\frac{e^2}{8\pi^2}\epsilon^{\mu\nu\rho\sigma}A_{\nu}\partial_{\rho}A_{\sigma}$
is the Chern-Simons current. It follows that the Chern-Simons number $n_{_{\rm CS}}=K^0$ is identified with the magnetic helicity density [we use a $(-,+,+,+)$ signature and $\epsilon^{0123}=-\epsilon_{0123}=1$]
 \begin{equation}
n_{_{\rm CS}}=\frac{e^2}{8\pi^2}\, \vec{A}\cdot\vec{B}\,.
 \end{equation}
In the particular case of a homogeneous fermion distribution, the anomaly equation reduces to (we fix $N_f = 1$ from now on)
 \begin{align}
\partial_0 J_5^0 & = \partial_\mu K^\mu\,.
\end{align}
This allows us to write an effective description without fermions.  Integrating them out, they can be represented by a (homogeneous) chemical potential $\mu$ sourced by the Chern-Simons number~\cite{Redlich:1984md,Niemi:1985ir}. \modif{Furthermore, in the case of massless fermions, the relation between the chiral current and the chemical potential can be written in a closed form as\footnote{This is obtained for the ensemble average of the number density with the Fermi-Dirac distribution in the presence of a chemical potential: $J_5^0 \equiv n_{+}-n_-$, with $n_{\pm} = {1\over 2\pi^2}\int_0^\infty dE\,E^2\left(1+e^{E \pm \mu \over T}\right)^{-1}$.}
\begin{equation}
  J_5^0 = \frac{1}{6}\mu T^2 +\frac{\mu^3}{6\pi^2}\,.
\end{equation}
}

The anomaly, which represents a violation of chirality,  \modif{can then be recast as an equation for $\mu$}
\begin{equation}
\frac{d}{dt}\left( \frac{1}{6}\mu T^2 +\frac{\mu^3}{6\pi^2} \right) = \frac{e^2}{8\pi^2} {1\over V}\int d^3x\,F_\mn \tilde{F}^\mn\,.
\label{mu5eq}
\end{equation}
The equations of motion of the scalar and gauge fields, together with the anomaly equation (\ref{mu5eq}), can be actually derived from an effective action
\modif{
\begin{align}
	S_{eff}= -\int d^4x\Big(&(D_\mu\phi)^*(D^\mu\phi) - V(\phi) + {1\over 4}F_\mn F^\mn -   \nonumber\\
    &{1\over 2}(\partial_0 a)^2 - {1\over 4 M^4}(\partial_0 a)^4  -  {e^2\over (4\pi)^2}{ a\over \Lambda}F_\mn \tilde{F}^\mn\Big)\,,
  \label{eq:actionGeneric}
\end{align}
}upon identifying $\partial_0 a= {\Lambda\mu}$, $\Lambda^2 = T^2/12$ and \modif{$M^4 = \frac{\pi^2T^4}{9}$}, where $T$ is the temperature of the system. The potential $V(\phi)$ was given in Eq.~\eqref{eq:potential}.

\modif{The main aim of this work is to extract the decay rate of the chemical potential in the presence of an external magnetic field. This quantity is related to Chern-Simons number diffusion rate in the absence of chemical potential through the fluctuation-dissipation theorem (we discuss this in detail in appendix \ref{app:fluc-diss}).  It is of special interest as the corresponding diffusion rate has been measured from (independent) simulations in Ref.~\cite{Figueroa:2017hun}. The fluctuation-dissipation relation is a near-equilibrium relation, relevant for small chemical potentials $\mu \ll T$. Having this in mind, we will restrict our attention to the reduced system}
\modif{
\begin{align}
	S_{eff}= -\int d^4x\Big((D_\mu\phi)^*(D^\mu\phi) - &V(\phi) + {1\over 4}F_\mn F^\mn - {1\over 2}(\partial_0 a)^2 -  {e^2\over (4\pi)^2}{ a\over \Lambda}F_\mn \tilde{F}^\mn\Big)\,,\nonumber\\
  \label{eq:action}
\end{align}
which corresponds to neglecting the $\mu^3$ term in the anomaly equation, hence leading to a linear dynamics in $\mu$. We have explored the dynamics of this system for different values of $\mu$, solving the linear equations of motion that follow from Eq.~(\ref{eq:action}). In the case of large chemical potential $\mu \gg T$, the correct physical regime is rather described by the non-canonical kinetic term of $a(t)$ that we have neglected in Eq~\eqref{eq:action}. Keeping that term as in the full action~(\ref{eq:actionGeneric}), would lead however to a non-linear set of equations of motion, requiring a more complicated algorithm for solving the evolution of the system in a lattice. In this work we focus for simplicity in the linearised description (at the level of the equations of motion) of the system given by Eq.~\eqref{eq:action}, independently whether this corresponds to the correct description of the physical regime expected for a given $\mu$. Fortunately, in order to extract the chiral decay rate of the chemical potential in the presence of a magnetic field, it is enough to explore the regime $\mu < T$, for which Eq.~(\ref{eq:action}) represents the correct description of the system.}

\modif{We start nonetheless studying the system with large non-vanishing initial values of the chemical potential, $\mu > T$. Even though, as mentioned, Eq.~\eqref{eq:action} does not correspond to the correct description of this physical regime, we still expect to get a qualitative understanding of the system dynamics. In this case, the chemical potential is expected to be unstable, forced to decay down to lower values. Due to volume effects unavoidable in any lattice simulation, the chemical potential cannot decay completely, and rather relaxes into a volume-dependent finite critical value $\mu_c$. By reaching a large enough volume in the lattice, we have achieved sufficiently small values of $\mu_c$, so that we can characterise well the chemical potential decay. In particular, we observe that the decay proceeds through a self-similar/power-law regime, while at the same time the gauge field long-wavelengths develop an inverse cascade spectrum.
}

\modif{Next, we move into the main aim of our present work, studying the system in the presence of a background magnetic field. In this case, the theory becomes similar to its non-Abelian counterpart, as the vacuum becomes degenerate. In particular, $\mu$ decays exponentially all the way down to zero. We have measured the corresponding rate $\Gamma_5 \equiv -{d\log \mu\over dt}$ in the regime of small $\mu \leq T$, where \eqref{eq:action} describes the physically relevant theory. We provide in this way a new determination of the parametric dependence of the chiral rate $\Gamma_5$, which exhibits a best fit scaling as $\Gamma_5 \propto e^{11/2}B^2$, with no residual volume dependence, where $B$ is the external magnetic strength. Furthermore, we have compared this prediction comparing the magnitude and parametric scaling of $\Gamma_5$ obtained from our present lattice simulations, against direct measurements of $\Gamma_{\rm diff}$ from (independent) simulations previously presented in Ref.~\cite{Figueroa:2017hun}, which according to the fluctuation-dissipation argument should be related as $\Gamma_5 = 6\Gamma_{\rm diff}/T^3$, with $T$ the temperature of the system.}

On the technical side, the discretisation of this theory needs to be done with care, especially when considering the Chern-Simons number. An appropriate discretisation scheme for Abelian gauge theories, reproducing the continuum limit of the theory to quadratic order in the lattice spacing, was presented in~\cite{Figueroa:2017qmv}. It obeys the following properties on the lattice: $i)$ the system is exactly gauge-invariant, and $ii)$ shift symmetry of the axion is exact\footnote{An equivalent formulation for non-Abelian gauge theories was originally introduced in~\cite{Moore:1996qs,Moore:1996wn}. In non-Abelian gauge theories the shift symmetry is however not preserved exactly on the lattice.}. Property $i)$ implies that physical constraints such as the Gauss law or Bianchi identities, are exactly verified on the lattice (up to machine precision). Property $ii)$ implies that the lattice formulation naturally admits a construction of the topological number density with a total (lattice) derivative representation
$\mathcal{K} \equiv F_\mn\tilde F^\mn =  \Delta_\mu^+ K^\mu$, which reproduces the continuum expression $\mathcal{K} = \partial_\mu K^\mu \propto \vec E \cdot \vec B$ up to $\mathcal{O}(dx_\mu^2)$ corrections. Without this property, the interaction $a\,F_\mn\tilde F^\mn$ cannot be interpreted as a derivative coupling, and hence is not really shift symmetric. As it is precisely the shift symmetry which justifies the functional form of the interaction in first place, it is therefore relevant to preserve exactly such symmetry at the lattice level. Hence, in the present work, we obtain results from  numerical simulations based on the discretisation scheme presented in Ref.~\cite{Figueroa:2017qmv}. In appendix, we provide a summary of the key equations of  such lattice formulation. For further technical details we refer the interested reader directly to Ref.~\cite{Figueroa:2017qmv}.

Most of what is known about anomalous $U(1)$ dynamics comes from MHD predictions, which represent the long-wavelength effective description of our system. \modif{To get analytic insights, it is interesting to compare our findings with this effective theory. The phenomena we observed (e.g.~self-similar decay of the chemical potential, gauge field inverse cascade dynamics, etc) are qualitatively well-modeled by MHD-inspired models, even though  these do not take into account the full-fledged short-scale dynamics present in our lattice simulations.}

The paper is organised as follows. In section 
\ref{sec:self-sim}, we  present the outcome of our lattice simulations. Namely, we study the evolution of an initially non-vanishing large chiral chemical potential $\mu$ in the absence of an external magnetic field. This leads to a characterisation of the self-similar decay of $\mu$, and of the gauge field inverse cascade dynamics. We then  add an external magnetic field, which introduce a vacuum degeneracy, and allows for the complete decay of the chemical potential. This gives us a way to measure \modif{in the physically relevant regime} the amplitude and parametric dependence of the decay rate $\Gamma_5$, which we compare with $\Gamma_{\rm diff}$ as inferred from previous simulations. Section \ref{sec:compMHD} is devoted to the comparison of our results to MHD.
We provide there a simplified MHD model which allows us to describe the gross feature of the system. We then characterise the inverse cascade phenomenon and see that its dynamical evolution can be well fit to an MHD-like $ansatz$. We also discuss the expected parametric dependence of the chiral rate.  In section \ref{sec:disc} we discuss our results and present future outlooks. Appendix \ref{app:latDesc} recaps the lattice setup and in appendix \ref{app:fluc-diss} we review the derivation of the fluctuation-dissipation relation between the chiral rate and the Chern-Simons diffusion one.

\section{Lattice results}
\label{sec:self-sim}

In this section, we report the results obtained from our lattice simulations. We select from a thermal ensemble (with $\mu=0$) initial configurations, which we then evolve following the (lattice version of) the classical equations of motions derived from \eqref{eq:action}, with initial condition $\mu(0)=\mu_0\neq 0$. We follow the evolution of the chemical potential, which in the symmetric phase is expected to be unconditionally unstable, leading to the creation of long-range gauge fields~\cite{Rubakov:1985nk,Rubakov:1986am}.  We start by investigating  the general features of the chemical potential evolution in the absence of a background magnetic field, making sure that the volume dependence of our observables is under control. Then, we study the self-similar behaviour of the decay for large initial chemical potential, and characterise the associated inverse cascade dynamics of the gauge field. We show that the magnetic power spectrum flows to the infrared modes. After that, we switch on a homogeneous background magnetic field. This is achieved by the use of twisted boundary conditions \cite{Kajantie:1998rz}, which are already introduced in the Monte-Carlo process that generates the initial configuration (we describe this in appendix \ref{app:latDesc}). We observe different regimes, and in particular, we focus on the exponential decay induced by the presence of the external magnetic field. This allows us to extract the chiral magnetic decay rate $\Gamma_5 = - {d\log \mu \over dt}$, and characterise its parametric dependence.

\subsection{Chemical potential decay}
\label{sec:chem_pot_finite_vol}

 The existence of an instability can be understood by looking at the free energy of the system under consideration. In momentum space, the usual magnetic term ${\mathcal H}_{\rm mag} = {1\over 2}{\vec B}^2 \sim {1\over 2}k^2 A^2$ competes with the chemical potential term ${\mathcal H}_{_{\rm CS}} = \mu n_{_{\rm CS}}\sim \pm \frac{e^2}{8 \pi^2}\mu k A^2$, where the different sign corresponds to the different polarisations of the gauge field, reflecting the chiral nature of the coupling. The chiral term may become dominant at small momenta, and for the gauge field polarisation for which ${\mathcal H}_{_{\rm CS}}$ and ${\mathcal H}_{\rm mag}$ have opposite sign, an instability occurs for sufficiently infrared modes
 \begin{equation}
 	k<\frac{e^2}{4 \pi^2}\mu =k_{c} .
	\label{eq:kmin}
 \end{equation}

\begin{figure}
	\centering
	\includegraphics{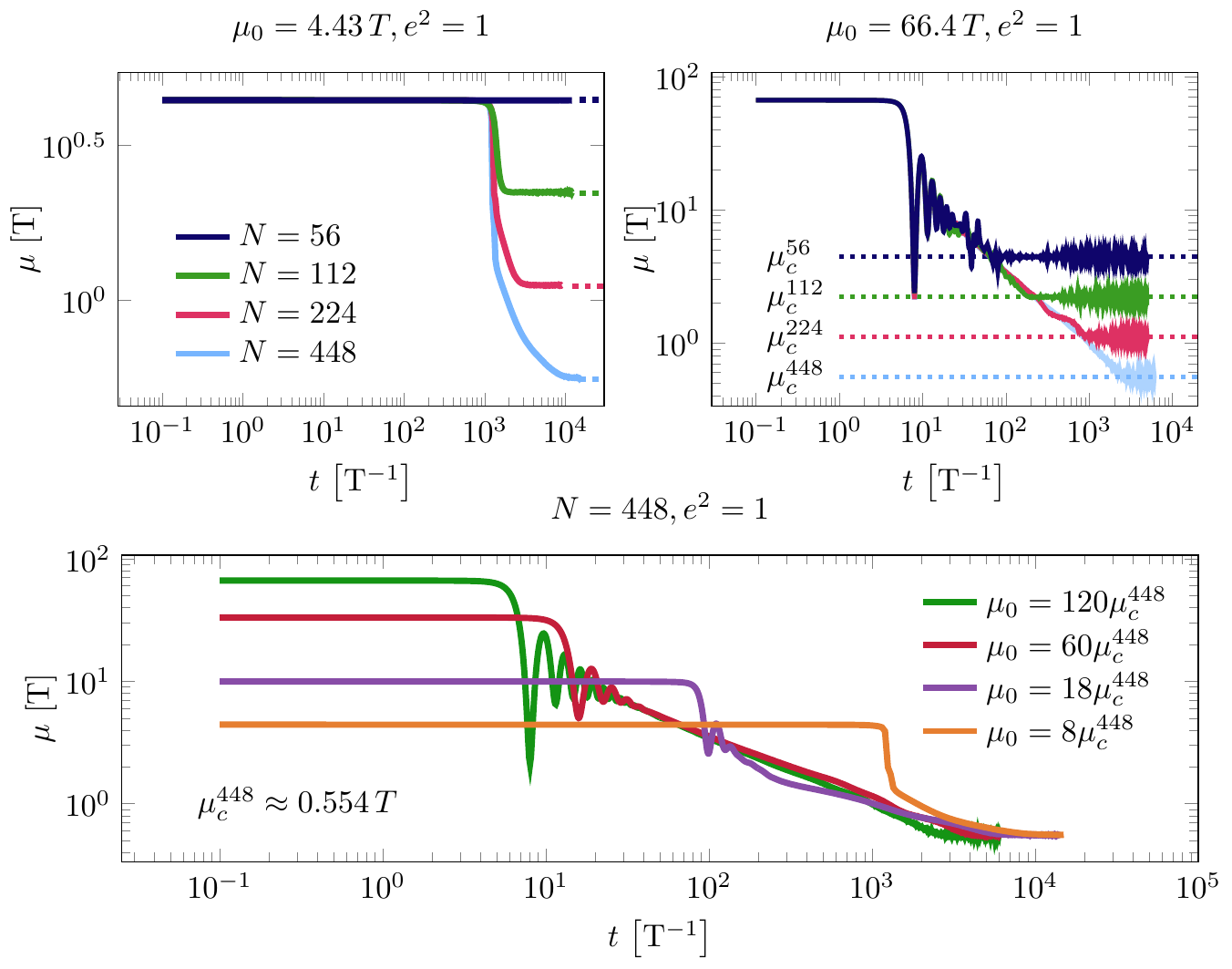}
	\caption{Chemical potential evolution for different lattice sizes and initial values. \textit{Upper panels:} Evolution of the chemical potential for different lattice sizes for a relatively small (left panel) and a relatively large (right panel) initial chemical potential. The dotted lines show the value of the critical potential predicted by equation \eqref{eq:mucrit}. As expected, finite volume effects are observed for chemical potential close enough to critical. Far enough from the critical value, all lattices give the same results. \textit{Lower panel:} Evolution of the chemical potential for different initial values, on the largest $N=448$ lattice we have simulated. As the initial value is decreased, the size of the initial $plateau$ increases. For large chemical potential, we observe a power law decay, which is related to some self-similar behaviour.}
	\label{fig:scan}
\end{figure}

In particular, if all $k$'s on the lattice are larger than the critical $k$, no instability can develop. As lattice momentum is discrete and has a minimum value $k_{min}$, it implies the existence of a critical chemical potential below which no instability can develop.
Equation \eqref{eq:kmin} can be rewritten to understand what is the largest chemical potential which is stable given a momentum resolution. Using the smallest momentum in a lattice $k_{min}=\frac{2\pi}{N\dd x}$, with $N$ the number of lattice points in one direction and $\dd x$ the lattice spacing, we find
\begin{equation}
	\mu_{c}=\frac{8 \pi^3}{e^2 N \dd x}\,.
	\label{eq:mucrit}
\end{equation}
For $\mu < \mu_c(e,N,\dd x)$ the instability is not captured any more for the given lattice. From now on we set $T\dd x=1$, as any re-scaling of the lattice spacing can be translated into an inverse  re-scaling of the gauge coupling constant $e^2$. In this way, relevant scales are always guaranteed to be captured in the lattice (see \cite{Figueroa:2017hun} for more details). In the upper panels of figure \ref{fig:scan} we show the behaviour of the chemical potential for different lattice sizes (in this figure and in the following, we indicate in brackets the corresponding units in powers of the temperature $T$). The critical threshold predicted by \eqref{eq:mucrit} is clearly appreciated, as indicated by the dotted lines in the figure. As long as the chemical potential remains much larger than the critical value for a given lattice size $N$, all simulations agree, as expected, independently of the volume. On the other hand, some finite volume effects are observed around the critical value. In particular, we see in the top right panel that for the smallest volume considered $N = 56$, the chemical potential displays some oscillations just before approaching the critical value; such oscillations disappear for larger volumes. In the lower panel, we display the chemical potential evolution for different initial values, for the largest $N=448$ lattices we simulated. For large enough initial values, we observe a power law decay, which as we will explain in section \ref{sec:inv_cascade}, corresponds to a self-similar behaviour.

Another phenomenon we observe is that the smaller the initial chemical potential, the longer it takes for the decay to be triggered. In other words, we see the emergence of long $plateau$s, gradually longer the smaller the initial value of the chemical potential is. This phenomenon makes the regime of very small chemical potential difficult to be captured well on a lattice. Atop of requiring large volumes to decrease the critical chemical potential, one has to perform increasingly longer simulations.

\subsection{Inverse cascade}
\label{sec:inv_cascade}

\begin{figure}[h]
	\centering
	\includegraphics{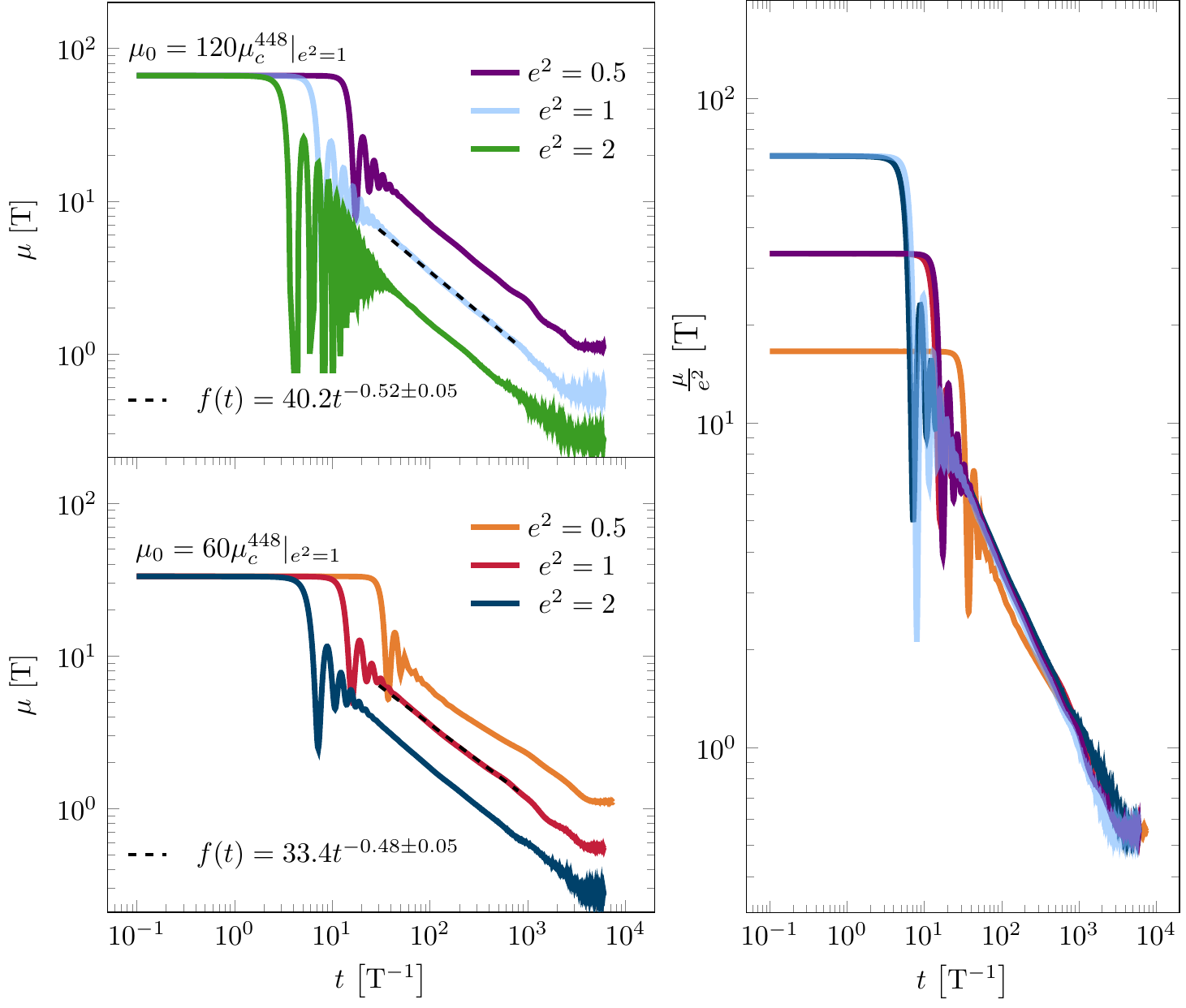}
	\caption{\textit{Left:} Chemical potential decay for different electric charges and different initial values. The dashed lines are the fits to a power-law (errors on the fit to the exponent are specified but not plotted). We see that the exponent is compatible with $-\frac{1}{2}$. \textit{Right:} Chemical potential normalised by $e^2$. The data collapse shows that $\mu(t)\propto e^2g(t)$ with $g(t)$ a function independent of $e^2$. }
	\label{fig:chem_decay}
\end{figure}

In the previous subsection, we have seen that finite volume effects are under control for sufficiently large lattices $N \gg 50$, and  that long initial $plateau$s appear for small chemical potentials. A remarkable feature of the chemical potential evolution in this context is that for large initial values, its decay corresponds to a universal power-law.
We show this phenomenon in figure \ref{fig:chem_decay} for different charges and initial values of the chemical potential. After some transient behaviour, all decays enter a power-law regime well described by a $t^{-1/2}$ behaviour. All the dependence on the parameters comes from the charge and lies in the prefactor as we illustrate in the right-hand side part of the figure.

\begin{figure}
	\centering
\includegraphics{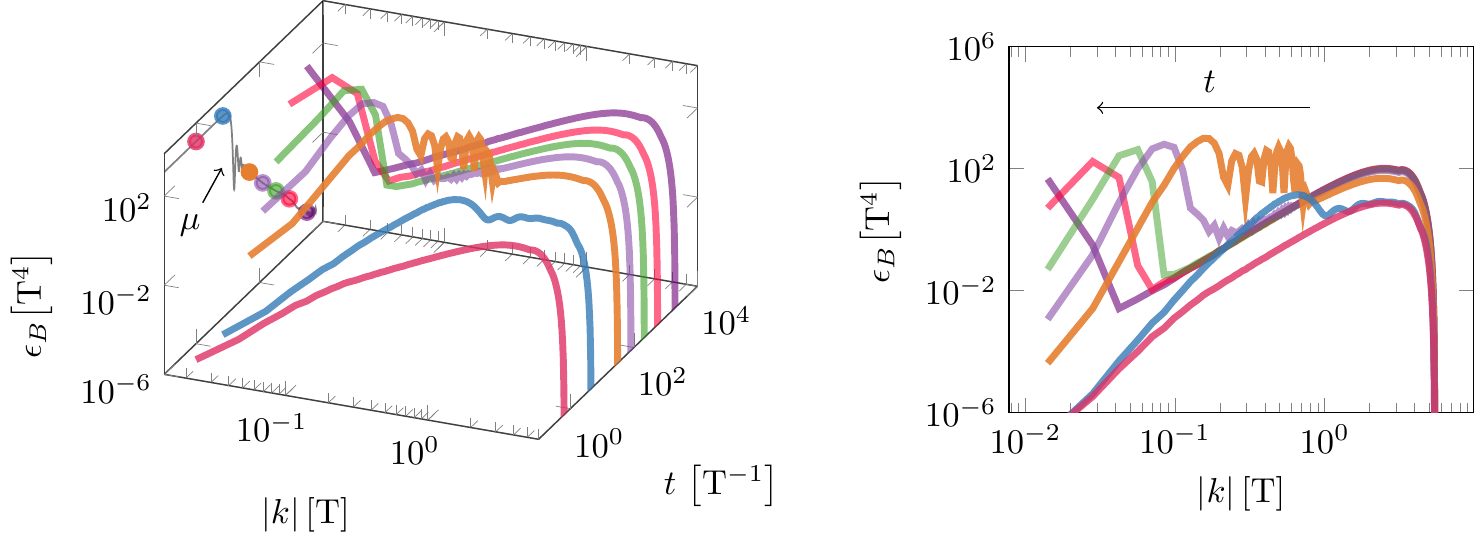}
	\caption{Magnetic power spectrum  on a $448^3$ lattice with $\mu_0=60\mu_c^{448}$ and $e^2=1$. In the $3D$-plot, we also show the evolution of the  chemical potential (the reader is referred to figure \ref{fig:chem_decay} for the correct scale). As the decay proceeds, we assist to a steady transfer of energy towards the IR. In the end, most of the magnetic  energy is stored in the minimal frequency. The UV part of the spectrum is related to the intrinsic UV sickness of the classical theory.}
	\label{fig:power-spectrum}
\end{figure}

Actually, this phenomenon goes in pair with what is known as an inverse magnetic cascade. The decay of the chemical potential induces a transfer of magnetic energy from the ultraviolet (UV) into the infrared (IR). In other words, long-range magnetic fields are created. We consider the magnetic power-spectrum
\begin{equation}
	\epsilon_B(k) = {1\over 2\pi^2} \left(\frac{\dd x}{N}\right)^3 \left\langle k^3|\tilde B(k)|^2\right\rangle_{|k|}\,,
  \label{eq:powerspectrumdef}
\end{equation}
where the quantity  $\langle\dots\rangle_{|k|}$ denotes an angular average over the spherical shell of radii $[|k|-{1\over 2}\Delta k,|k| + {1\over 2}\Delta k)$, with $\Delta k = k_{\rm min} \equiv {2\pi\over N dx}$ the binning width.
\modif{The quantity $\tilde {B}(k) = \sum_x e^{\vec{-ik}\cdot\vec{x}} \vec{B}(x)$ is the lattice discrete Fourier transform of $\vec{B}(x)$ and has the same units as $\vec{B}(x)$.}  This definition has the advantage of being volume independent and it is simply related to the real-space volume average as
 $ \langle {\vec B}^2\rangle_{_{V}} = \sum_{k} \frac{k_{\rm min}}{k}\,\epsilon_B(k)$
 mimicking the continuum relation $\langle {\vec B}^2\rangle = \int\frac{\dd k}{k} \,\epsilon_B(k)$.

\begin{figure}
\centering
\includegraphics{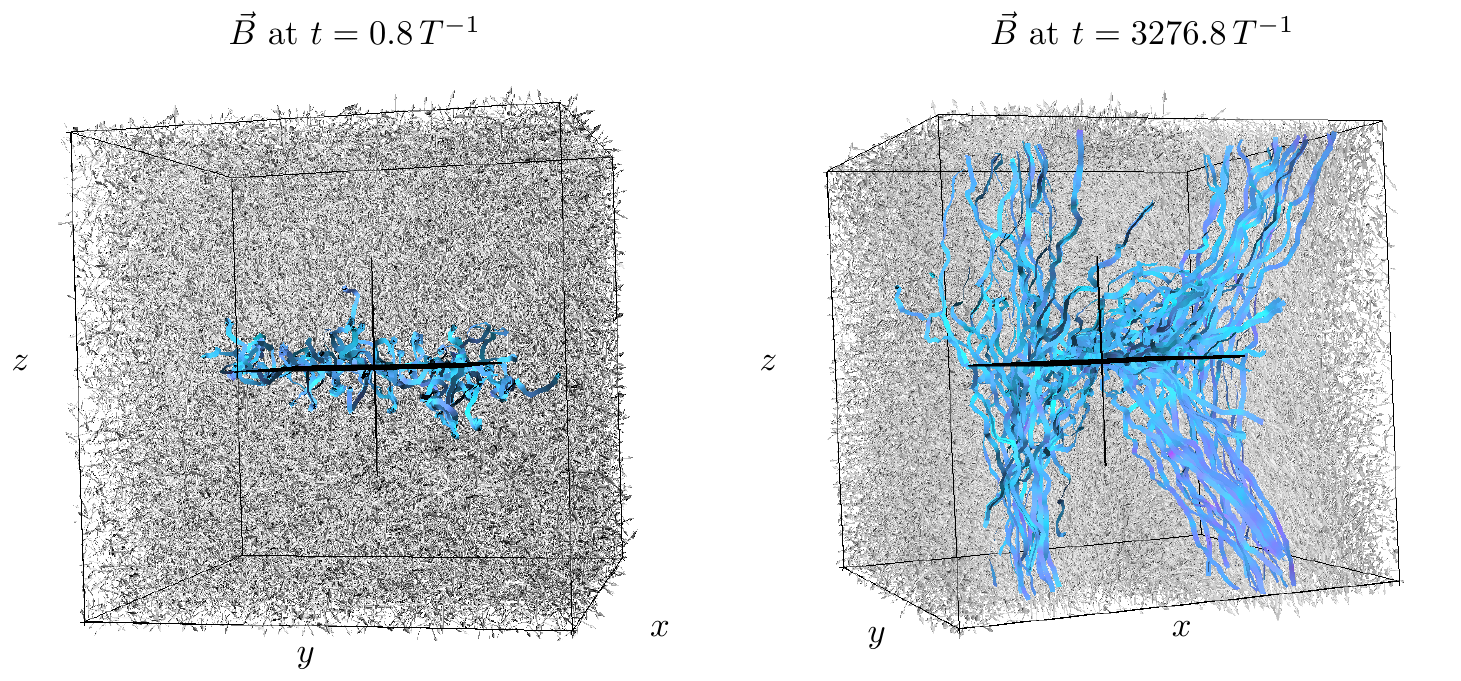}
\caption{Snapshots of the magnetic field just after thermalisation (left panel) and at the end of the inverse cascade (right panel) on a $448^3$ lattice with $\mu_0=60\mu_c^{448}$ and $e^2=1$. The blue 'ribbons' are field lines emanating from a small plane placed in the center of the box. They can, for instance, be drawn by following the motion of a test magnetic charge initially on the plane. At early times, no special structure is seen, since the magnetic field is homogeneous and well thermalised. At the end of the simulation, after the inverse cascade process has operated for a long while, the magnetic field has developed a long-range order, leading to structures which permeate the whole lattice. This is the spatial counterpart to the IR power displacement in Fourier space described below Eq.~(\ref{fig:power-spectrum}).}
	\label{fig:spatial_B}
\end{figure}

In figure~\ref{fig:power-spectrum}, we plot the time evolution of the magnetic spectrum, where an energy flow from UV scales into IR scales, i.e.~an inverse cascade, is clearly observed. As the chemical potential starts decaying, the magnetic energy is gradually transferred into the lower modes, so that at the end of the simulation most of it is peaked around the smallest lattice mode $k_{\rm min}$. 
Another way to display this information is by looking directly at the spatial distribution of the magnetic field, as we do in figure~\ref{fig:spatial_B}. There we show snapshots of the magnetic field arrows, together with some representative field lines emanating from the center. In the left panel of figure~\ref{fig:spatial_B}, we see that, immediately after thermalisation, no characteristic structures in the magnetic field distribution are observed. However, after the inverse cascade process occurs, we see long-range magnetic fields, as depicted in the right panel of figure~\ref{fig:spatial_B}, where the field lines reach out through the whole lattice.

\subsection{Decay in the presence of an external magnetic field}
\label{sec:ov_dyn_B}
The dynamics of a chiral charge is of special interest in the presence of a background magnetic field $B_i=\frac{1}{2}\epsilon_{ijk}F_{jk}$. In this case, as detailed in \cite{Figueroa:2017hun}, the vacuum of the theory is degenerate and the situation is closely related to its non-abelian counterpart. As already mentioned, such a background can be introduced on the lattice through the use of twisted boundary conditions \cite{Kajantie:1998rz}. The main difference from the case without magnetic field is that the chiral charge is now unconditionally unstable, and can decay all the way down to zero.

The early dynamics of the system depends crucially on the strength of the magnetic field. For large enough ones, it will drive the chiral charge decay from the beginning onward. For weak external magnetic fields, we expect the system to evolve initially like in the absence of magnetic field, and settle down to the critical value $\mu_c$. The chemical potential cannot stay however in the state $\mu=\mu_c$ in the presence of a background magnetic field, and eventually  decays to zero.

\begin{figure}
	\centering
	\includegraphics{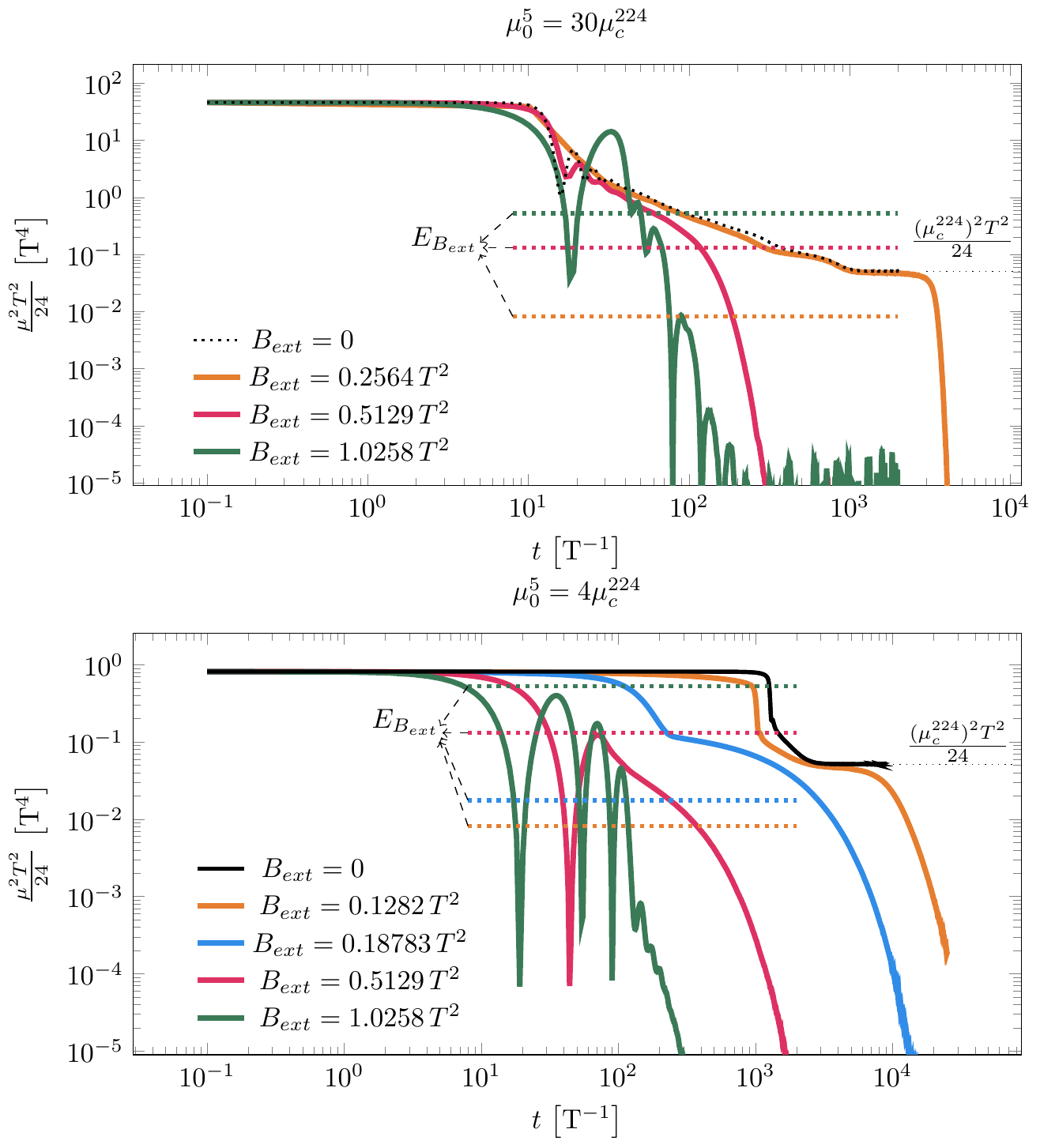}
	\caption{Chiral chemical potential energy evolution for a variety of external magnetic fields. \modif{ The lines $E_{B_{ext}}=\frac{1}{2}B_{ext}^2$ are the energies of the external magnetic field, to give a reference point.} \textit{Upper plot:} Large initial chemical potential. We see three regimes. First, when the external magnetic field's energy is much smaller than the initial chemical potential energy, it has no effect on the initial dynamics; the system evolves as if there was no external background (orange curve). After some time spent  in the critical $plateau$, the system eventually decays to a $\mu=0$ state. For a large external magnetic energy, the system does not enter the self-similar regime and the decay happens through  the external magnetic field from the beginning (green curve).  Between these two regimes, there is an intermediate one, where both effects contribute (pink curve). \textit{Lower plot:} Small initial chemical potential. A similar discussion applies as in the upper plot, though the system exhibits now higher sensitivity to the external magnetic field, as the ratio between the chemical and magnetic energy is smaller. } 
	\label{fig:mag_dep}
\end{figure}

These two limiting situations can be well observed  in our simulations. We can also study the transition between them, see in particular figure \ref{fig:mag_dep}. There we plot the contribution to the total energy from the chemical potential \modif{$\frac{\mu^2 T^2}{24}$}\footnote{\modif{Of course this represents a "physical" energy only for $\mu<T$.} }. This allows us to compare the chemical potential energy with respect to that of the external magnetic field. In the upper panel, we plot configurations with large initial values of the chemical potential, whereas the analogous plots for weaker initial chemical potential values are presented in the lower panel. As a reference, we also display the outcome from simulations with the same initial chemical potential but no magnetic field.
The dashed lines represent the magnetic energy carried by the external background. Looking first at the weakest magnetic fields, we see that its influence on the chemical potential is minimal. For larger initial chemical potentials, the simulations are almost not affected while in the case of smaller chemical potentials the length of the initial $plateau$ is slightly reduced. As expected, we see that the chemical potential relaxes first to its critical value, before undergoing a secondary decay due to the presence of the background magnetic field.
For stronger background magnetic fields (top red  curve, and bottom red and blue curves), we observe a transition regime where both effects are competing. In this case, the system is not sensitive any more to the finite number of infrared modes; the chemical potential does not stabilise to a critical value. In both cases, it happens when the external magnetic energy is roughly two orders of magnitude smaller than the initial chemical potential energy. For even larger  magnetic energies, we enter into a different dynamical regime, where the chemical potential decays quickly to zero through damped oscillations.

\begin{figure}
	\centering
	\includegraphics{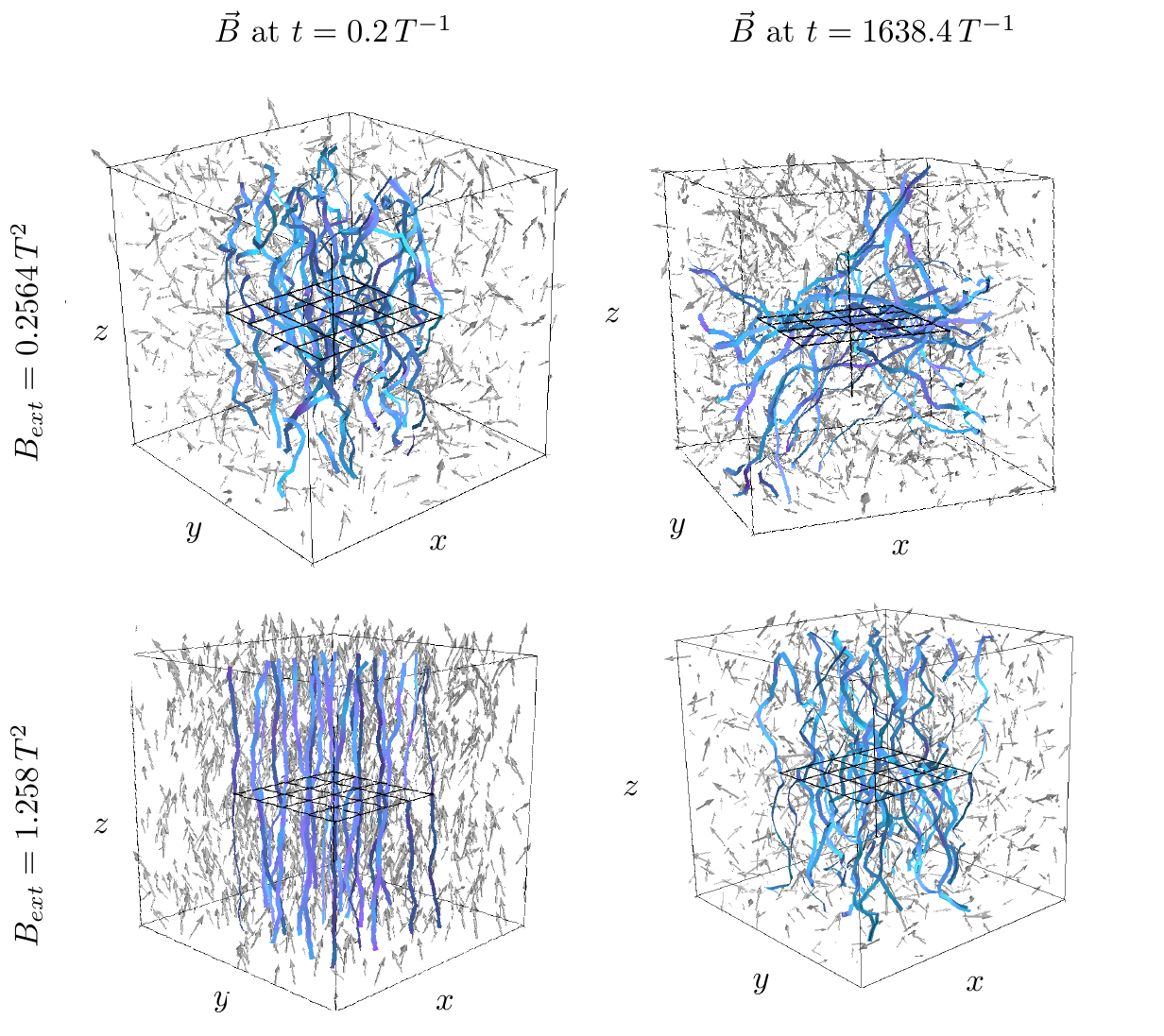}
	\caption{Magnetic fields after thermalisation (left panels) and at the end of the evolution (right panels), obtained for $N=224$ and $\mu^{0}_5=30\mu_c^{224}$. Blue ribbons are the field lines emanating from a subplane in the center of the lattice. For both values of the external magnetic field, the initial configuration looks similar, with the magnetic field lines oriented along the $z$-direction. The final states are however different. For the weaker $B_{\rm ext}$ (orange curve of figure \ref{fig:mag_dep}), the magnetic field develops long-range structures in transverse directions to the $z$-direction. For the larger $B_{\rm ext}$, the final structures tend to be more aligned to the $z$-direction.}
	\label{fig:ext_mag_visu}
\end{figure}

It is also instructive to look at the spatial distribution of the magnetic field. In figure~\ref{fig:ext_mag_visu}, we show the magnetic field  lines corresponding to the orange and red curves in the top panel of  Fig.~\ref{fig:mag_dep}, associated with large initial chemical potentials. In the left panels of Fig.~\ref{fig:ext_mag_visu}, we show the field lines just after thermalisation, and in the right panels we show the field line distribution at the final stage of the decay.  As expected, the magnetic field is initially oriented along the direction of the external magnetic field (which, without loss of generatily, was chosen to be oriented along the $z$-axis), and the stronger the external magnetic field the less important the thermal fluctuations are. In the final stage, we still see lattice-size structures. For weaker magnetic fields, the decay is essentially driven by the chiral instability of the chemical potential, so the field lines extend in all directions, as in the case without magnetic fields. Yet the effect it is still visible due to the presence of the external magnetic field, as more field lines expand along the $z$-direction. For a larger magnetic field background, the field lines which are not aligned to the $z$-axis are suppressed at the end of decay, and we essentially see only field strength lines along the $z$-axis.

\subsection{Chiral magnetic rate}
\label{subsec:RateComparison}
As presented in the previous section, the evolution of the system in the presence of an external
magnetic field is subject to two competing dynamics. On the one hand, there is the intricate evolution related to the anomaly-induced creation of long-range gauge fields. On the other hand, the external magnetic field acts as a vacuum reservoir and induce an exponential decay of the chemical potential. In this section, we are interested in this dynamic.

\begin{figure}
	\centering
	\includegraphics{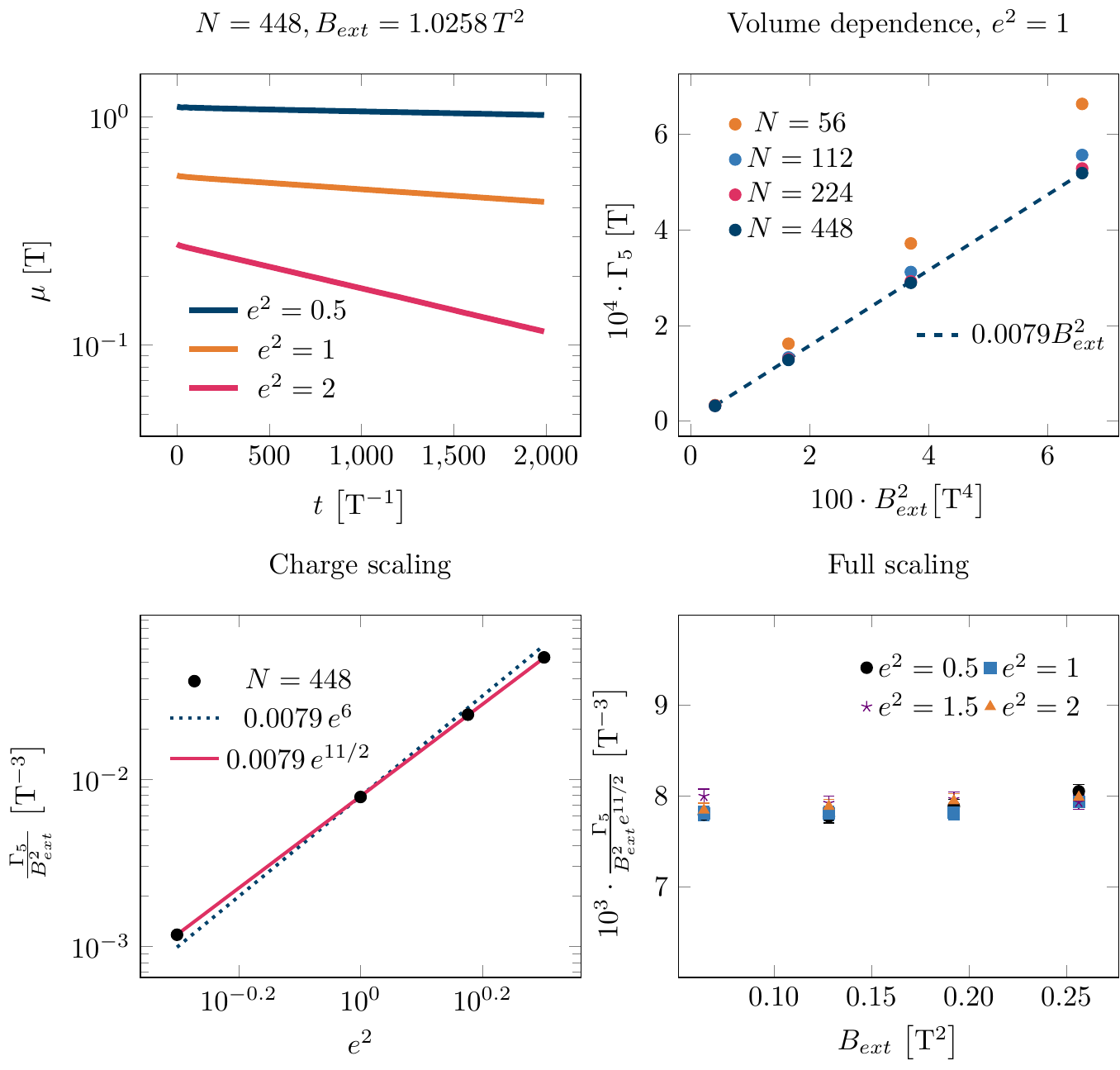}
	\caption{\textit{Upper left:} Exponential decay of the chemical potential for $\mu_{5}^0=\mu_c$. \textit{Upper right:} Volume dependence of the measured decay. We observe convergence to in the infinite volume limit. The $N=448$ results differ only in a few parts in thousands from a naive linear extrapolation using the $N=224$ and $N=448$ results. \textit{Lower left:} Charge dependence of the decay rate. We observe a deviation from the dominant $e^6$ scaling. We can well describe it by an effective $e^{11/2}$ dependence. \textit{Lower right:} Full scaling of the rate. We see that the quantity $\frac{\Gamma_5}{e^{11/2}B_{ext}^2}$ is almost constant. The remaining charge and magnetic field dependence is very weak.}
	\label{fig:gamma_rate}
\end{figure}

To extract $\Gamma_5$ from our simulations, we proceed as follow. To disentangle the magnetic driven decay from the rest of the dynamics, we take our initial chemical potential to be the critical one. \modif{Thanks to our large volumes, it also allow us to reach the physically relevant regime $\mu<T$.} As explained in section~\ref{sec:chem_pot_finite_vol}, the anomaly-related instability does not develop for sub-critical chemical potentials. Representative evolution of the critical chemical potential in the presence of an external field are shown in the upper-left plot of~\ref{fig:gamma_rate}. After an initial relaxation, we see, as expected, a clear exponential decay. This allows us to measure $\Gamma_5$. We show its $B^2$ dependence for different volumes in the upper-right panel of~\ref{fig:gamma_rate}. We observe a good convergence to the infinite volume limit, as the $N = 448$ and $N = 224$ cases differ only in $\sim 0.5\%$. We thus pursue the investigation only on the $N=448$ lattices, adding a $1\%$
systematic error to take into account any potential remaining volume dependence. We obtain that there is virtually no deviation from the expected $B^2$ dependence.

In the lower-left panel of Fig.~\ref{fig:gamma_rate}, we look at the charge dependence of the rate. Fitting our data, we find that they are well modeled by a $\propto e^{11/2}$ dependence. This can also be seen in the lower-right panel of Fig.~\ref{fig:gamma_rate}, where we show $\frac{\Gamma_5}{e^{11/2}B_{ext}^2}$. All the different determinations collapse to a constant; there is very little remaining dependence on the charge and magnetic field.

More quantitatively, we present in Table~\ref{tab:table1} the different fits we performed to our data. All the main fits agree with each other, giving an almost exact $B^2$ magnetic dependence, an effective $e^{11/2}$ charge dependence and a prefactor of $10^{-2.1}\approx 0.0079$.
To check the robustness of the coefficients, we also to tried to fit $\frac{\Gamma_5}{e^6}$ with $f(B)=\alpha B^\beta$. Such a hypothesis is clearly excluded by the data as it leads to  a $\chi^2/\mathrm{dof}$ of $155$. We also present in this table a fit in $\Gamma_5= a B^2 e^6 \ln\left(\frac{b}{e}\right)$ to show that the charge dependence can also potentially interpreted as logarithmic corrections to a leading $e^6$ scaling. As we will discuss in section \ref{subsec:MHD_rates}, this is what can be expected from theoretical predictions.

\begin{table}
  \centering
\begin{tabular}{c|c|c}
  Fit quantity & Fit  & $\chi^2/\mathrm{dof}$\\
  \hline
  $\Gamma_5$ & $10^{-2.094\pm 0.008}(B)^{2.001\pm0.009}(e^2)^{2.76\pm0.01}$ & $0.88$ \\
  $\frac{\Gamma_5}{B^2}$ & $10^{-2.103\pm 0.002}(e^2)^{2.75\pm0.01}$ & $1.13$ \\
  $\frac{\Gamma_5}{e^{11/2}}$ & $10^{-2.098\pm 0.008}(B)^{2.005\pm0.008}$ & $1.13$ \\
  $\frac{\Gamma_5}{e^{6}}$ & $10^{-2.0\pm 0.1}(B)^{2.1\pm0.1}$ & $155$ \\
  $\frac{\Gamma_5}{e^{11/2}B^2}$ & $10^{-2.10\pm0.02}$ & $1.16$\\
  $\frac{\Gamma_5}{e^6 B^2}$ & $(1.9\pm 0.1)\cdot 10^{-3}\ln\left(\frac{62\pm 13}{e^2}\right)$ & $1.80$\\
  \hline
\end{tabular}
\caption{Chiral dissipation rate $\Gamma_5$. The errors are two standard deviations and are not rescaled by the $\chi^2/dof$. }
\label{tab:table1}
\end{table}

These numbers may be compared to the values of the topological charge diffusion rate, which at large time reads
\begin{equation}
  \langle Q^2(t)\rangle =\Gamma_{\rm diff} V t
  \label{eq:diff_rate}
\end{equation}
with $Q(t)=N_{cs}(t)-N_{cs}(0)$, $N_{\rm cs}(t)=\frac{e^2}{8\pi^2}\int \dd x^3\vec{A}\cdot \vec{B} $. This quantity was measured in  \cite{Figueroa:2017hun}, and it is related to $\Gamma_5$ through a standard fluctuation-dissipation argument\footnote{Note that the corresponding  expression in~\cite{Figueroa:2017hun} contains a factor $2$ wrong, see appendix \ref{app:fluc-diss} here, for a clarification of this.}~
\begin{equation}\label{eq:FluctDiss}
  \Gamma_5=6 \frac{\Gamma_{\rm diff}}{T^3}.
\end{equation}

\begin{table}
  \centering
 \begin{tabular}{c|c|c}
   Fit quantity & Fit  & $\chi^2/\mathrm{dof}$\\
   \hline
   ${ \left\langle Q^2 \right\rangle \over Vt}$ & $10^{-3.00 \pm 0.23}\cdot (e^2)^{2.87\pm 0.12} \cdot B^{2.06\pm 0.07}\cdot V^{0.05 \pm 0.09}$&$0.67$\\
${ \left\langle Q^2 \right\rangle \over Vt e^6} $& $10^{-3.06 \pm 0.24} \cdot 10^{-3}\cdot B^{2.05\pm 0.08}\cdot V^{0.06 \pm 0.09} $&$0.74$\\
${ \left\langle Q^2 \right\rangle \over Vt e^{11/2}} $& $ 10^{-2.93 \pm 0.24} \cdot 10^{-3}\cdot B^{2.06\pm 0.08}\cdot V^{0.03 \pm 0.06}$ &$0.74$\\
${ \left\langle Q^2 \right\rangle \over Vt B^2}$ & $10^{-2.91 \pm 0.21} \cdot 10^{-3}\cdot (e^2)^{2.87 \pm 0.12}\cdot V^{0.03\pm0.04}$& $0.70$\\
${ \left\langle Q^2 \right\rangle \over Vt e^6 B^2}$ & $10^{-2.90\pm 0.10}\cdot V^{0.06\pm 0.08}$ &$0.76$\\
${ \left\langle Q^2 \right\rangle \over Vt e^{11/2} B^2}$ & $10^{-2.84\pm 0.20}\cdot V^{0.00\pm 0.04}$ &$0.77$\\
\hline
 \end{tabular}
 \caption{Diffusion rate $\Gamma_{\rm diff}$ from \cite{Figueroa:2017hun}, re-analysed. We take advantage of this re-analysis to take into account the statistical fluctuations in the fit, which was not done in \cite{Figueroa:2017hun}. The errors represent two standard deviations. The relatively small $\chi^2/dof$ shows some degree of overfitting. As we do not use it to rescale the errors, they are probably overestimated.}
 \label{tab:table2}
\end{table}

For the sake of the comparison, we present in Table~\ref{tab:table2} a re-analysis of the data of Ref.~\cite{Figueroa:2017hun}. Unfortunately, the diffusion rate is a harder quantity to extract and leads to a less precise quantification. As the results of ~\cite{Figueroa:2017hun} were obtained at different small volumes, we keep an explicit volume dependence in our fits, even if we find it to be weak. Taking as a measured value $\Gamma_5= 10^{-2.10\pm 0.02} B^2 e^{11/2}/T^3$ leads to a prediction
 $\Gamma_{\rm diff}=10^{-2.88\pm0.02} B^2 e^{11/2}$. The prefactor and the magnetic field dependence are consistent with the measurements of $\Gamma_{\rm diff}$ reported in~\cite{Figueroa:2017hun}. At first sight, the scaling with $e^2$ seems to be in mild tension, but further analysis shows that the data from Ref.~\cite{Figueroa:2017hun} does simply not constrain well enough the charge dependence of $\Gamma_{\rm diff}$. This can be seen in the fits in Table~\ref{tab:table2}, where either scalings $\propto e^{11/2}$ and $\propto e^{6}$ are compatible with the data, exhibiting a similar $\chi^2/dof$. We conclude that within the errors, the fluctuation-dissipation relation Eq.~(\ref{eq:FluctDiss}) is well verified.

\section{Comparison to MHD}
\label{sec:compMHD}

\modif{Different approaches can be followed to get some analytical understanding of a system in the presence of a chiral chemical potential. Recently, progresses have been made towards establishing a $chiral-$kinetic theory, which adds to the usual Boltzmann-like approach a self-consitent treatment of the anomaly, see \cite{Chen:2015gta} and references therein for more information. Valuable knowledge may also be gained in the hydrodynamical regime by studying the effective field theory of the long wavelength field modes together with the electromagnetic interactions. In this approach, matter is a fluid,  characterised by its velocity field. A variety of interaction terms going from the simple linear response electric conductivity to direct coupling of the velocity which can induce turbulence can be added. This kind of approach is referred to as MHD.  The equations can be further modified to take into account anomalous processes \cite{Giovannini:1997eg,Joyce:1997uy}, to give anomalous MHD. Recently, an extensive study was carried out, both at the theoretical level \cite{Rogachevskii:2017uyc} and using numerical simulations, \cite{Schober:2017cdw} of full anomalous MHD models. We refer the interested reader to these references for more details. }

\modif{To gain some understanding of our results,we will attempt to model the system using a truncated set of MHD equations. Namely, we will focus on the magneto-dynamics part, not considering any hydrodynamics evolution.  }

\begin{eqnarray}
\begin{array}{lcl}
\derp{\vec{E}}{t} = e\vec{j} - \vec{\nabla}\times \vec{B} -\frac{e^2}{4\pi^2}\mu\vec{B}\,, & \hspace*{1cm}& 	\derp{\vec{B}}{t} = \vec{\nabla}\times \vec{E}\,, \vspace*{0.3cm}\\
\frac{\dd \mu}{\dd t}=\frac{3 e^2}{T^2\pi^2}\frac{1}{V}\int \dd x^3\vec{E}\cdot\vec{B}\,, &\hspace*{1cm}& 	\vec{j}=-\sigma \vec{E}\,.
\end{array}
\label{eq:MHD}
\end{eqnarray}
where $\sigma$ is the linear-response electric conductivity.

\modif{This approximation will turn out to qualitatively work surprisingly well. Indeed, as our lattice simulations describes the full-dynamics of our matter field, there is no $a priori$ reasons for the matter field to contribute little to the global dynamics. It is even likely that in some regimes not explored in this work, matter effects which would be described as turbulence in a hydrodynamic language should manifest themselves.}

\modif{In particular, this model predicts the existence of a chiral magnetic rate. Indeed, the evolution equation for $\mu$ can be rewritten as, neglecting the time derivative of the electric field,}

\modif{\begin{align}
  \frac{\dd \mu}{\dd t}=-\frac{3 e^2 }{T^2 \pi^2 \sigma}\frac{1}{V} \int \dd x^3 \left ( \frac{e^2}{4 \pi^2}\mu\vec{B}+\vec{\nabla}\times \vec{B}\right)\vec{B}\,.
\end{align}
Taking $B$ to be constant, we then obtain
\begin{align}
  \mu\propto e^{-\Gamma_5^{(MHD)} t},
\end{align}
with $\Gamma_5^{(MHD)}=\frac{3 e^4 }{4 \pi^4 \sigma T^2}$. We will compare it to our results in section \ref{subsec:MHD_rates}.}

\modif{Finally, note that from now on, when we refer to MHD or MHD-like predictions, we have in mind the specific model \eqref{eq:MHD}.}

\subsection{Qualitative behaviour and initial plateaus}
\label{subsec:plateau}

As presented in previous sections, the chemical potential is transferred into long-range helical magnetic fields which carry a non-vanishing Chern-Simons number. To understand at a qualitative level some of the observed features, like the existence of $plateau$s in the initial stage, we can consider the evolution of a maximally helical field  under anomalous MHD-like dynamics. In the presence of a background magnetic field in the $z$- direction, this corresponds making the $ansatz$ \cite{Rubakov:1985nk}

\begin{equation}
	A_1=f(t)\sin(kz)-\frac{B}{2}y, \ \ A_2=f(t)\cos(kz)+\frac{B}{2}x, \ \ A_3=f_z(t)\,,
	\label{eq:ansatz2}
\end{equation}
which leads to

\begin{align}
	\ddot{f}&=kf\left (\frac{e^2\mu}{4\pi^2}-k \right )-\sigma \dot{f},\label{eqAp:pheno_model1}\\
	\dot{\mu}&=\frac{3e^2}{T^2\pi^2} \left (B\dot{f_z}-kf\dot{f}\right ),\\
	\ddot{f_z} &= -\sigma\dot{f_z} -\frac{e^2}{4\pi^2}\mu B.
\end{align}
Let us first consider this toy-model without an external magnetic field ($B=0, f_z(t)=0$). Then the system of equations reduces to
\begin{align}
	\ddot{f} =kf\left (\frac{e^2\mu}{4\pi^2}-k \right )-\sigma \dot{f}\,,~~~~~~
  \mu=-k\frac{3e^2}{2T^2\pi^2}f^2+k\frac{3e^2}{2T^2\pi^2}f_0^2+\mu_0\,.
\label{eq:pheno_model}
\end{align}

\begin{figure}
\centering
\includegraphics{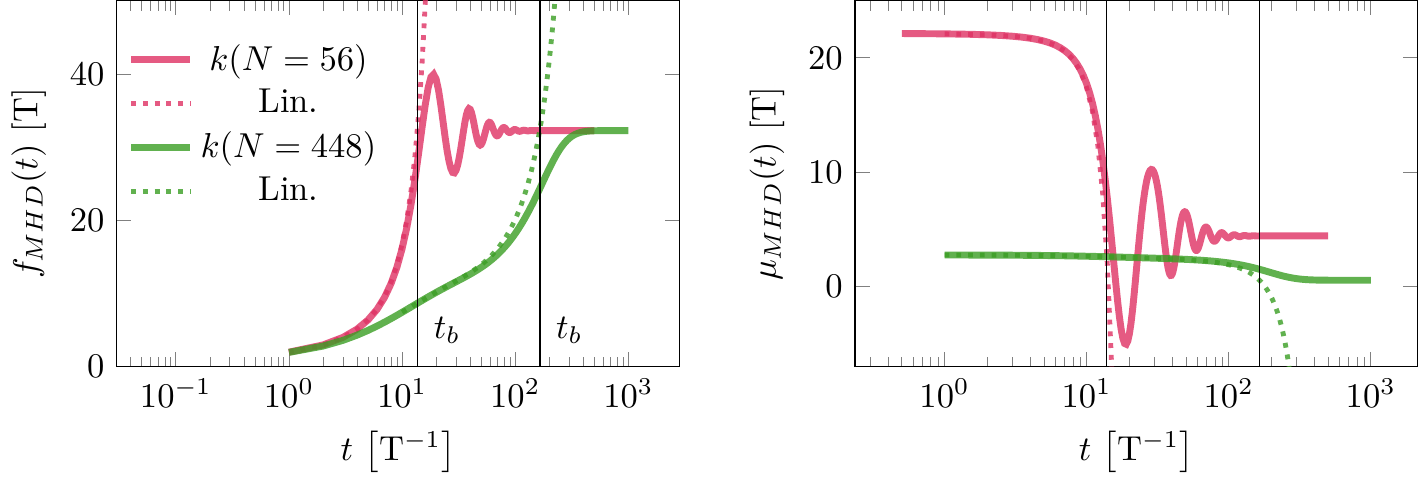}
\caption{\textit{Left:} Single-mode MHD solution for some typical value of the parameters. The value of $k$ is chosen to correspond to the minimal $k$ which can be simulated on lattices of sizes $N=56$ and $N=448$. We further set $\sigma=0.1, e^2=1$ and take $f_0=\dot{f_0}=1$ as initial values. We also plot the predicted breaking time, which gives a good estimate for the end of the exponential growth. \textit{Right:} Corresponding chemical potentials. }
\label{fig:pheno_model}
\end{figure}
\begin{figure}
	\centering
	\includegraphics{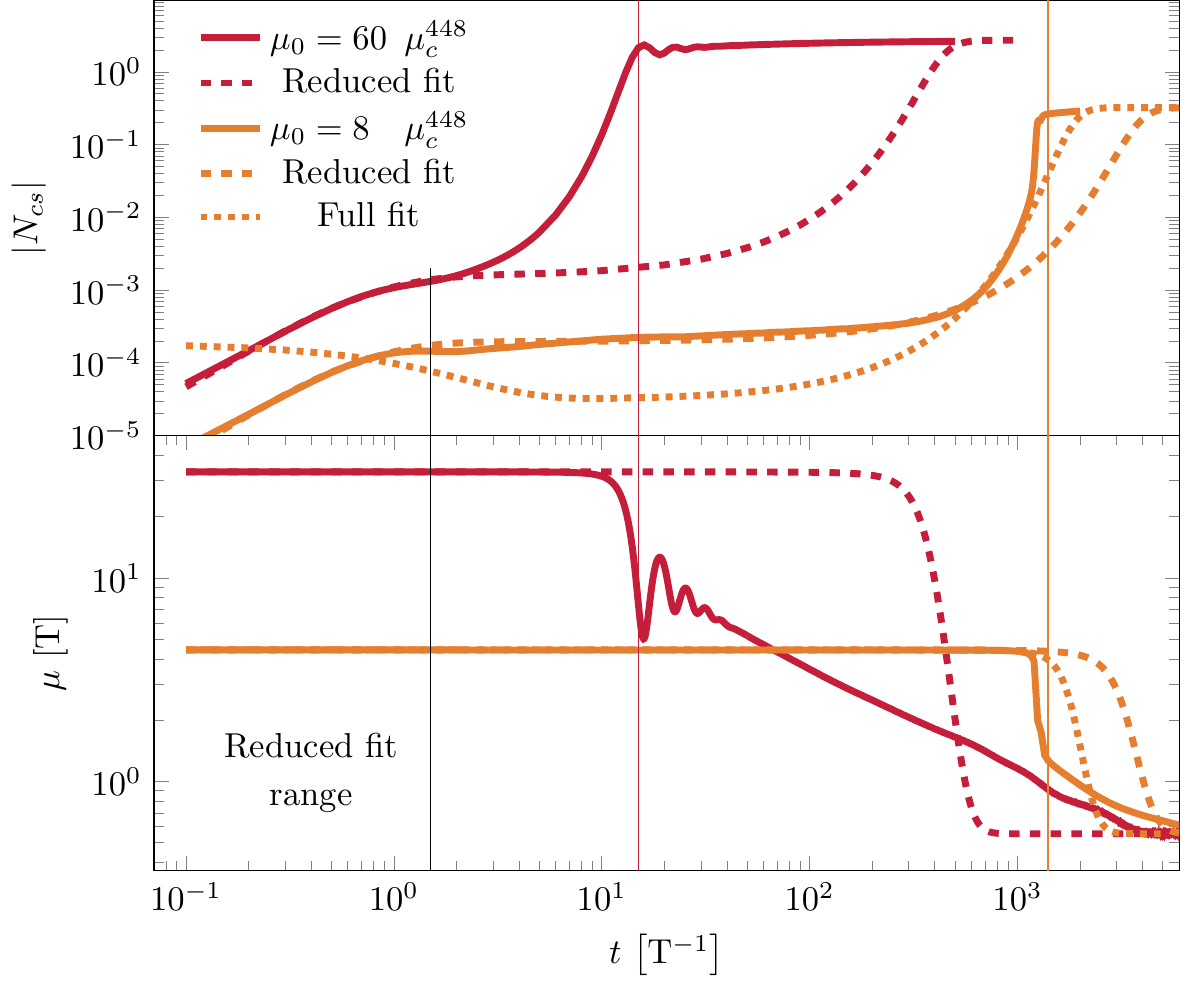}
	\caption{\textit{Upper panel:} Chern-Simons number evolution. \textit{Lower panel:} Chemical potential. In both cases, non-solid lines are fits obtained from the model \eqref{eq:pheno_model}. Dotted ones correspond to a fit over the full time range whilst dashed lines were obtained by restricting $t<10$. We see that the model is appropriate only for the initial part of the evolution.  }
	\label{fig:chiral}
\end{figure}

By plugging in the expression for $\mu$ into the first equation of (\ref{eq:pheno_model}), we obtain a non-linear equation for $f(t)$. Solving it numerically also gives us access to the evolution of $\mu(t)$. \modif{Here, we take $f_0$ and $\sigma$ to be some free parameters to be fitted.} Examples of the time evolution of $f(t)$ and $\mu(t)$ obtained with this procedure, are shown in figure~\ref{fig:pheno_model}.
Despite being in qualitative agreement, this simple modeling does not capture completely the fine details, especially for large initial chemical potentials. In figure \ref{fig:chiral}, we compare the modeling based on the {\it ansatz} \eqref{eq:ansatz2} with the numerical outcome from our lattice simulations. The lower panel shows the chemical potential whilst the upper panel shows the Chern-Simons number, which is useful to understand the initial dynamics. Solid lines were obtained from our simulations, while dashed and dotted ones are numerical solutions to equations~(\ref{eq:pheno_model}) for different parameters (conductivity and initial conditions).

Let us first consider the case $\mu_0 = 8\mu_c$, as the model gives a better description for relatively small chemical potentials.
The dotted curves were obtained by fitting the $ansatz$ model to the whole range of data. When looking at the chemical potential, we see that the model is able to describe reasonably well the initial $plateau$, but fails to describe well the decay.  Actually, when looking at the Chern-Simons number, we see that even the initial phase is not very well described by this fit. A remedy to this is to restrict the fitting range to early times, where the system should be closer to the model. For example, the dashed curves are obtained by restricting the fitting range to times $t < 10/T$ (delimited by a solid vertical black line). This gives a much better description of the initial phase; essentially the initial evolution is well captured by the simple $ansatz$. At a later stage, when a more complex dynamics has developed, such as the creation of long-range magnetic fields following a process of inverse cascade (or what can be described as a turbulent regime in MHD), the model is too simplistic to capture well the physics. The same can be said about large initial chemical potentials, as in the case of $\mu_0 = 60\mu_c$ depicted in the figure. There, we see explicitly that the power-law decay (expected due to the generation of long-range magnetic field) is completely missed by the $ansatz$ modeling. The red and orange vertical solid lines are a visual guide indicating the saturation of the Chern-Simons number and of the corresponding chemical potential.

We can get a better understanding of the early dynamics of $f(t)$ and $\mu(t)$ in the  $ansatz$ modeling, by simply perturbing  equation~(\ref{eq:pheno_model}), i.e.~considering $f(t)=f_0+\delta f(t)$, so that
\begin{align}
	f_0 k^2 -\frac{e^2}{4\pi^2}f_0 k\mu_0 +k^2 \delta f(t) +\frac{3e^4}{4\pi^4 T^2}f_0^2 k^2\delta f(t) -\frac{e^2}{4\pi^2}k\mu_0 \delta f(t)+\sigma \dot{\delta f}(t)+\ddot{\delta f}(t) \notag\\
	+\frac{9 f_0 k^2e^4}{8 \pi^4 T^2}\delta f(t)^2 +\frac{3  k^2e^4}{8 \pi^4 T^2}\delta f(t)^3=0.
  \label{eq:linearisation_MHD}
\end{align}
 Keeping the linear order terms in $\delta f$ and neglecting the terms proportional to $f_0^2$ (we do not start in a state with an helical field), the equation reduces to
\begin{equation}
		\left (k^2 -\frac{e^2}{4\pi^2}k\mu_0\right )\left(f_0 +\delta f(t)\right)+\sigma \dot{\delta f}(t)+\ddot{\delta f}(t)=0\,.
\end{equation}
This is a driven harmonic oscillator which admits a solution as
\begin{equation}\label{eq:pertSOL}
\delta f(t)\approx	\frac{e^{-\frac{\sigma}{2}t}}{2\omega_{eff}}\left (2f_0\omega_{eff} \cosh\left(\omega_{eff}t\right)+(2 \dot{f_0}+f_0\sigma)\sinh\left(\omega_{eff}t\right) \right )-f_0\,,
\end{equation}
where $\dot{f_0}$ is the initial time derivative of $f(t)$, and we have defined
\begin{equation}
	\omega_{eff}^2=\frac{e^2}{4\pi^2}k\mu_0-k^2+\left(\frac{\sigma}{2}\right )^2\,.
\end{equation}
Eq.~(\ref{eq:pertSOL}) indicates, first of all, that the solution corresponds to an IR instability for the modes $k<\frac{e^2}{4\pi^2}\mu_0$, which grow exponentially fast. Secondly, by estimating the range of validity of this solution, we can estimate the duration of the chemical potential $plateau$ until the onset of its decay. Indeed, the breakdown of this approximation corresponds to the end of the exponential growth of $f(t)$, which in turn triggers the chemical potential decay. The approximations cease to be valid when higher order perturbations become non-negligible. To see this in detail, we keep  only the exponentially growing part of the solution and neglect the constant term,
\begin{equation}
\delta f(t)_{as}\approx	\frac{\exp\left(\left(-\frac{\sigma}{2}+\omega_{eff}\right )t\right )}{2\omega_{eff}}\left (f_0\left(\omega_{eff}+\left ( \frac{\sigma}{2}\right )\right ) + \dot{f_0} \right )\,.
\end{equation}

The relative weight between the linear to second-order perturbation terms in the equation of motion \ref{eq:linearisation_MHD}, and linear to third order terms, are defined by the ratios
\begin{align}
r_2(t)=\frac{k^2   -\frac{e^2}{4\pi^2}k\mu_0}{\frac{9 f_0 k^2e^4}{8 \pi^4 T^2}}\frac{1}{\delta f(t)_{as}}\,, ~~~~~ r_3(t)=\frac{k^2   -\frac{e^2}{4\pi^2}k\mu_0}{\frac{3  k^2e^4}{8 \pi^4 T^2}}\frac{1}{\delta f(t)_{as}^2}
\end{align}
 We expect the linear approximation to breakdown whenever either of these ratios become of order one. We thus define the breakdown time $t_b$ to be
 \begin{equation}
t_b\equiv \min(t_2,t_3)\,,~~~~~~{\rm with}~~	r_2(t_2) \equiv 1\,,~r_3(t_3) \equiv 1\,.
	\label{eq:tb}
\end{equation}
In figure \ref{fig:pheno_model}, we show how this prediction performs, comparing the linearisation (dotted lines) to the numerical solution to Eq.~(\ref{eq:pheno_model}). We see that $t_b$ in~\eqref{eq:tb} gives a reliable prediction of the range of validity of the linearisation regime. In conclusion, the duration of the initial $plateau$ can be estimated well with our simple MHD-inspired $ansatz$ equation~\eqref{eq:ansatz2}.

\begin{figure}
	\centering
	\includegraphics{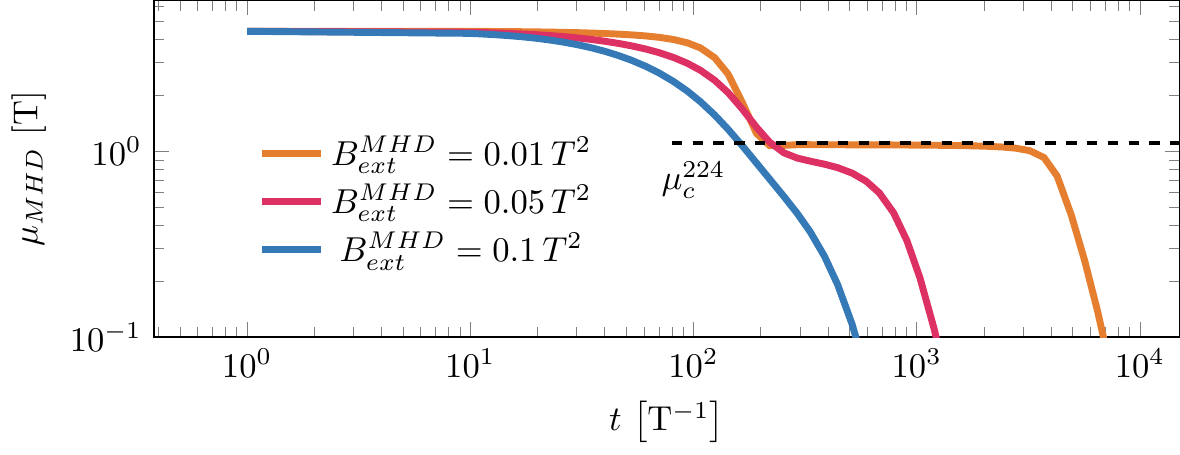}
\caption{Numerical solution to our single-mode MHD model for three different magnetic fields and some reasonable initial conditions. The behaviour is in qualitative agreement with the numerical outcome from lattice simulations, see Fig.~\ref{fig:mag_dep}. Three regimes appear to be present: one initially dominated by the chiral instability (orange curve), another where the decay is strongly induced by the external magnetic field (blue curve), and an intermediate regime (pink curve).}
\label{fig:pheno_model_B}
\end{figure}

In the presence of an external magnetic field, the system of equations (\ref{eqAp:pheno_model1}) does not admit a simple enough analytic treatment. We can solve them nonetheless numerically. We show the chemical potential obtained from numerical integration of these equations in figure \ref{fig:pheno_model_B}. As in the previous case, we see that it captures reasonably well the dynamics of small chemical potentials. For weak magnetic fields, it is dominated by the chiral instability of the chemical potential. The system evolves as in the absence of magnetic fields, reaches a $plateau$ and only at later times it decays to zero because of the presence of the magnetic background field. For a sufficiently large background magnetic field, the decay is driven by the presence of such field, driving quickly the system into a $\mu=0$ state. For intermediate external magnetic fields, we have a superposition of the two effects.

\subsection{Self-similarity and inverse cascade}
\label{subsec:MHD_inv_casc}

The power law decay $\propto t^{-1/2}$ of the chemical potential reported in section \ref{sec:inv_cascade} was also predicted in the context of MHD in several works~\cite{Joyce:1997uy,Boyarsky:2011uy,Boyarsky:2015faa,Tashiro:2012mf,Hirono:2015rla}.
More specifically, various analysis of the time evolution of the magnetic spectrum, have been carried out in refs.~\cite{Joyce:1997uy,Tashiro:2012mf,Hirono:2015rla}. We will now compare these modelings with the outcome from our numerical simulations. First of all, we repeat the analysis of refs.~\cite{Joyce:1997uy,Tashiro:2012mf,Hirono:2015rla}. The starting point in MHD are equations ~\eqref{eq:MHD}. Taking the curl of the first equation and neglecting the second time derivative of $ \frac{\partial^2 \vec{B}}{\partial t^2}\approx 0$, the equation for the magnetic field can be recast as
\begin{equation}
	\sigma \frac{\partial \vec{B}}{\partial t} = \nabla^2 \vec{B} +\frac{e^2}{4 \pi^2}\mu(\vec{\nabla} \times \vec{B})\,.
  \label{eq:MHD_B_only}
\end{equation}
We can now decompose the magnetic field in an orthonormal helicity basis $Q^{\pm}(\vec{k},\vec{x})$, which corresponds to a basis\footnote{The precise form of the basis depends on the base space the analysis is performed, e.g. Ref.~\cite{Joyce:1997uy,Tashiro:2012mf} considers $\mathbb{R}^3$ whilst Ref.~\cite{Hirono:2015rla} used a sphere. In the analytical description above we will simply work in a Euclidean base space $\mathbb{R}^3$, as we are describing the continuum theory. Were we to write down a basis for the torus, we could construct it nonetheless from a superposition of the basis of~\cite{Tashiro:2012mf}.} of eigenvectors of the curl operator,
\begin{equation}
	\vec{B}(x,t)=\int \frac{\dd^3 k}{(2\pi)^3}\left (\vec{\alpha}^+(t,\vec{k}) Q^+(\vec{k},\vec{x})+\vec{\alpha}^-(t,\vec{k}) Q^-(\vec{k},\vec{x})\right )\,,
	\label{eq:spec_eq}
\end{equation}
so that $\vec{k}\times Q_{\pm}(\vec k) = \pm k Q_{\pm}(\vec k)$. In that  basis Eq.~(\ref{eq:MHD_B_only}) reduces to
\begin{equation}
	\frac{\partial \alpha^\pm(t,\vec{k})}{\partial t}=\frac{1}{\sigma}\left ( -k^2\pm\frac{e^2}{4 \pi^2}\mu(t) k \right)\alpha^\pm(t,\vec{k})\,,
\end{equation}
which admits as a solution
\begin{equation}
	\alpha^\pm(t,\vec{k})=\alpha^\pm_0(\vec{k})\exp\left [\frac{1}{\sigma}\left ( -k^2t\pm k\frac{e^2}{4 \pi^2}\int_{t_0}^t\dd t'\mu(t') \right)\right ]\,.
	\label{eq:sol_mode}
\end{equation}
Furthermore, as soon as $\int_{t_0}^t\dd t'\mu(t')$ becomes sizable, $\alpha^-(t,\vec{k})$ becomes sub-dominant. Neglecting the latter, we can then write
\begin{equation}
	|B(t,\vec{k})|^2\approx|\alpha^+_0(\vec{k})|^2 \exp\left [\frac{2}{\sigma}\left ( -k^2t + k\frac{e^2}{4 \pi^2}\int_{t_0}^t\dd t'\mu(t') \right)\right ]\,.
	\label{eq:pred_spec}
\end{equation}

\begin{figure}
	\centering
	\includegraphics{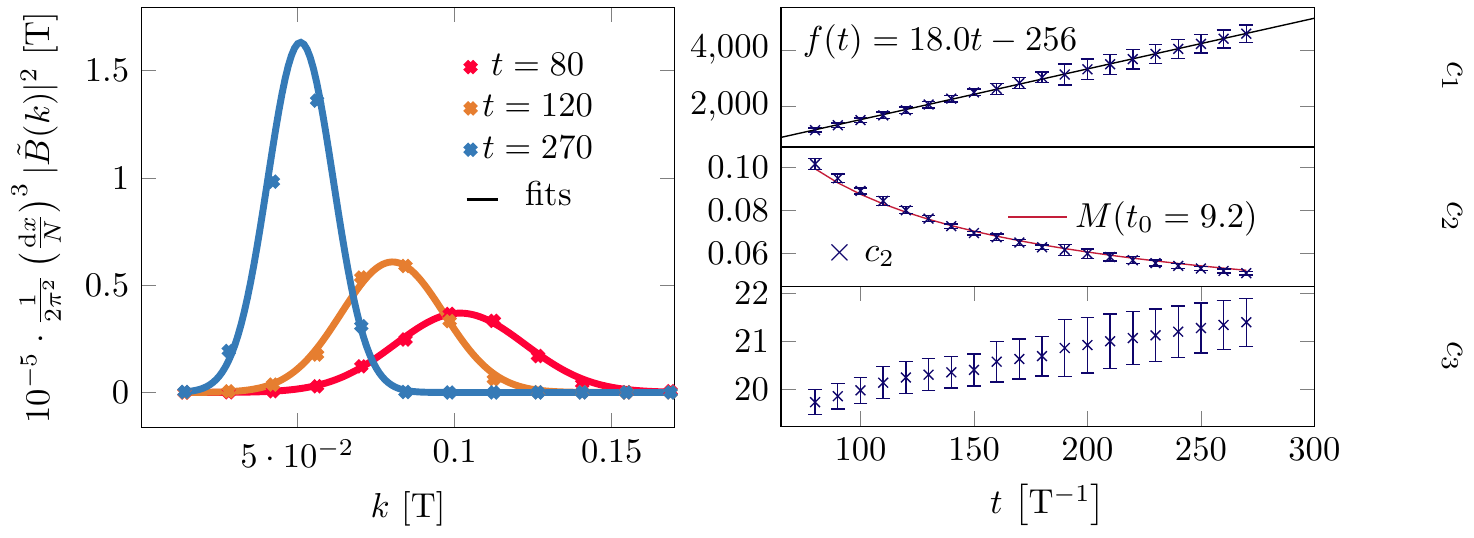}
	\caption{\textit{Left:} Time evolution of \modif{ $\frac{1}{2\pi^2}\left(\frac{\dd x}{N}\right)^3|\tilde{B}(k)|^2$ against MHD inspired fits, see equation \eqref{eq:powerspectrumdef}}. \textit{Right:} Fit's parameters as a function of time. The coefficient $c_1$ allows for a determination of an effective conductivity, $\sigma_{eff}$, $c_2$ can be used to check the validity of the fit, as it is expected to follow the evolution of the integrated chemical potential, which is indeed the case. Lastly, $c_3$ can also be used as \textit{a posteriori} check, as it is supposed to be a constant. The fits are based on MHD predictions and show a remarkable agreement in the qualitative description.}
	\label{fig:sig_fit}
\end{figure}

Equation \eqref{eq:pred_spec} represents a prediction based on MHD, which we can use to compare against our data. A further simplification we make, following \cite{Tashiro:2012mf}, is to neglect the $\vec{k}$ dependence\footnote{Note that we could also get rid of it by considering ratios of $|B(t,\vec{k})|^2$ at different times. An analysis using this method turned out to be less accurate, we believe because of the small number of momenta we can use for the fits.} of the initial amplitude $\alpha^+(\vec{k})\approx \alpha_0^+$. This leads us to write down the following function
\begin{equation}
B_{fit}^2(k,c_1,c_2,c_3)=\exp\left(c_1 k^2 +c_1\cdot c_2 k + c_3\right)\,,
	\label{eq:fit_model}
\end{equation}
which will be used as a fit to $|B(k,t)|^2$ obtained from our simulated data at different times. This is precisely what we show in figure~\ref{fig:sig_fit}. In the left panel we plot $|B(k,t)|^2$ for three different times together with a best fit of the form \eqref{eq:fit_model}. Despite the limited number of modes available, we see that the data is well fit by a functional form as in Eq.~(\ref{eq:fit_model}).

Let us look now at the time dependence of the fit coefficients $c_1, c_2$ and $c_3$. From comparing \eqref{eq:fit_model} and \eqref{eq:pred_spec}, we expect $c_1$ to be a linear function of $t$. This is exactly what we observe in figure \ref{fig:sig_fit}, where the top right panel displays a linear fit from our data as $c_1(t) = (18.0\pm 0.1)t + (256\pm 20)$. The crossing of the fit with the y-axis encodes actually various effects. First of all, we learn that $t_0$ is not zero, and rather corresponds to an initial time at which $|B(t,\vec{k})|^2$ is equal to $|\alpha^+_0| ^2$
(we will estimate it shortly, in a self-consistent manner, from the fit to the second coefficient $c_2$). 
Secondly, a non-zero crossing may
also reintroduce some $k^2$ dependence in the initial condition, which will partially compensate for the approximation ~$\alpha^+(\vec{k})\approx \alpha_0^+$.

\modif{ The coefficient $c_2$ is expected to go as $M(t_0,t)\equiv\frac{1}{t}\frac{e^2}{4 \pi^2}\int_{t_0}^t\dd t'\mu(t')$. Its dependence on $t_0$ allows us to estimate this initial time, by comparing it to $M(t_0,t)$ for different values of $t_0$, and looking for the best match. This is shown in the middle right panel of figure \ref{fig:sig_fit}. This procedure leads to an estimated value of as $t_0\approx 9.2/T$. A last check is provided by the lower right panel of figure \ref{fig:sig_fit}, which shows how close $c_3$ is to a constant.
The fact that this MHD-inspired model achieve such a good description of the data is a bit surprising, as the chemical potential probed here lie outside of the range of validity of \eqref{eq:MHD}. The relevant length scales which are to be compared are the typical instability length scale associated to the chemical potential decay, $l_{inst}\sim \frac{1}{k_{inst}} = \frac{4\pi^2}{e^2 \mu}$ to the typical mean free path in the plasma $\lambda\sim\frac{1}{\alpha^2 T}$. MHD is expected to be valid for $l_{inst} > \lambda$, which happens for $\mu < e^2 T$. The only place where the "breaking" of the effective theory can be seen is if we try to extract a conductivity from the coefficient $c_1$.
 This gives an effective conductivity $\sigma_{eff} = (0.11\pm 0.01) T$, which differs from the kinetic theory prediction, see section \ref{subsec:MHD_rates} for more details.
}

\modif{An interesting question would be to compare $\sigma_{eff}$  to the  intrinsic measurement of our system's conductivity with the use of  the Kubo formula~\cite{Kubo:1957mj}.  For classical field theories on a lattice with finite volume $V$, for a homogeneous and isotropic plasma,  $\sigma$ can be written as~\cite{Figueroa:2017hun}
\begin{eqnarray}
 \sigma 
 = \int dt \,\Sigma(t) ~,~~~~
 \Sigma(t) \equiv \frac{1}{V T}\frac{1}{3}\sum_{i=1}^3  \int d^3 x \int d^3 y \left\langle  j_i(x,t) j_i(y,0)\right\rangle,
\label{eq:integrand}
\end{eqnarray}
where $j_i$ are the spatial components of the electric current, and $ \langle ...\rangle$ is an ensemble average over different realisations of our thermal initial conditions. This formula is readily to be implemented as an observable in our lattice simulations. The expression is however UV sensitive (in the free theory it is even linearly divergent with momentum), so the results are expected to be sensitive to the natural UV cutoff imposed by the lattice $k_{\max} \sim 1/dx$. The correlator in (\ref{eq:integrand}) is expected to decay exponentially in time, $\Sigma(t) \sim \exp{(-\gamma t)}$, and to oscillate with the plasma frequency. When obtaining numerically $\Sigma(t)$ from our simulations, it exhibits an oscillatory pattern with frequency considerably smaller than $1/T$, which is presumably attributed to the plasma frequency (plots of this can be found in figure 7 (right panel) of Ref.~\cite{Figueroa:2017hun}, so we do not reproduce them here again). After some time, oscillations occur around zero, indicating the dumping. However,  the presence of short time oscillations, associated with the lattice UV cutoff, is also very noticeable. These contribute to make the behaviour of $\Sigma(t)$ very  'noisy',
 making difficult to obtain a trustworthy attempt to extract $\sigma$ by this procedure.
}

Note that even if we were to extract a conductivity from this method, there is no reason to expect it to be equal to the quantum field theory computation. Indeed, as it is a UV-dominated quantity, it is sensitive to the distribution of large-momentum modes, which are intrinsically different between the classical and quantum field theory, see \cite{Arnold:1997yb} for a careful discussion on this issue.

\subsection{Chiral magnetic rate and MHD}
\label{subsec:MHD_rates}

As mentioned at the beginning of the section, one can infer a chiral magnetic-rate from MHD
\begin{equation}
\Gamma_5^{(MHD)}=\frac{3 e^4 }{4 \pi^4 \sigma T^2}.
\label{eq:chiral_mag_rate}
\end{equation}
 \modif{To be able to use this prediction, we need to use  the kinetic theory  prediction for the conductivity.  The computation consists in finding the coefficient $\sigma$ in the relation $\vec{j} = -\sigma \vec{E}$ where $\vec{E}$ is an external uniform and time-independent electric field and $\vec{j}$ is the induced electric current. The main process to be taken into account is the mutual scattering of charged particles with an energy of the order of temperature. At leading-log order, for one charged fermion species, it can be estimated as~\cite{Arnold:2000dr}
\begin{eqnarray}
\sigma_{\rm F} \simeq {12^4\zeta(3)^2\over \pi^3(3\pi^2+32)}\frac{T}{e^2\log(q_{\rm F}/e)}\,.
\label{eq:cond_Fermion}
\end{eqnarray}
For one charged scalar species, the conductivity has been also obtained recently as~\cite{Sobol:2019kiq}
\begin{eqnarray}
\sigma_{\rm S} \simeq \frac{3^22^5\zeta(3)^2}{\pi^3}\frac{T}{e^2\log(q_{\rm S}/e)}~\,.
\label{eq:cond_Scalar}
\end{eqnarray}
The coefficients $q_{\rm F}, q_{\rm S}$ are $\sim \mathcal{O}(1)$, but can only be fixed from matching a computation at full leading order. The leading behaviour of the chiral rate is then expected to scale, independently of whether we use the fermion or scalar computation, as
\begin{align}
  \Gamma_5 \propto e^6 B^2\ln(q/e)/T^3\,,
\end{align}
with $q$ some $ \mathcal{O}(1)$ parameter.}

\modif{In the case of a charged fermion species, the coefficient $q_{\rm F}$ in the log can be determined using the full leading numerical result, for instance evaluated at $e^2=\frac{4\pi}{137.04}$, as can be read out from table 2 of Ref.~\cite{Arnold:2003zc}. Matching such value to Eq.~(\ref{eq:cond_Fermion}), leads to $q_{\rm F} = 4.2$, and hence to
\begin{eqnarray}
\sigma_{\rm F} \simeq {12^4\zeta(3)^2\over \pi^3(3\pi^2+32)}{T\over e^2\ln(4.2/e)}~.
\label{eq:cond_MHD}
\end{eqnarray}
Unfortunately, there is no analogous full leading computation available for the scalar field case, so we cannot determine exactly the value of $q_{\rm S}$. Assuming that $q_{\rm S} = q_{\rm F}$, then $\sigma_{\rm S}$ remains $\sim 14.5\%$ smaller than $\sigma_{\rm F}$. If we consider, however, values within the range e.g.~$q_{\rm F} - 2 \leq q_{\rm S} \leq q_{\rm F} + 2$, then $\sigma_{\rm S}$ ranges, for $e = 1$, between $\sim 56\%$ larger and $\sim 33\%$ smaller, compared to $\sigma_{\rm F}$ given by Eq~(\ref{eq:cond_MHD}) evaluated at $e = 1$. The difference becomes however smaller for smaller values of $e$.
}

In section \ref{subsec:RateComparison},  \modif{ were we extracted this rate in the physical region $\mu<T$ ,}  we observed an almost perfect  $B^2$ dependence of $\Gamma_5$, but a certain deviation from an exact $\propto e^6$ scaling.  In particular, fitting to a simple power law $\propto e^p$, we obtain a best fit with $p=11/2$, whereas fitting to a form $e^6\ln\frac{q}{e}$ we obtain $q=7.9$, see Table~\ref{tab:table1}\footnote{Note that we present there the result of the logarithmic fit as a function of $q^2\approx 62$.}.
 Both of these fits describe well our data, although the logarithmic fit leads to a slightly worse $\chi^2/dof$.  Given our limited range of values of $e^2$, we cannot clearly discriminate between these two options.

 \modif{Using as a reference $\sigma_F$, we can now estimate a chiral magnetic decay rate from MHD and compare it to our simulations. }
 Plugging in numbers, we get
\begin{eqnarray}
\label{eq:MHDcond}
\sigma_{\rm F} \simeq {31.4 \over e^2\log({17.6/e^2})}\,T~,
\end{eqnarray}
which for $e^2 = 1$ gives $\sigma_{\rm F} \simeq 10.95\, T$.  In the presence of a background magnetic field $B$, the chiral rate is actually inversely proportional to the conductivity, and it is given by
\begin{eqnarray}\label{eq:Gamma5MHD}
 \Gamma_5^{(MHD)} = \frac{3e^4}{4\pi^4\sigma_{F} T^2}\,B^2 \simeq 2.45\cdot 10^{-4}\log({17.6/e^2})\,e^6B^2 /T^3,
\end{eqnarray}
which for $e^2 = 1$ is $\Gamma_{5}^{(MHD)}(e^2 = 1) \simeq 7.0\cdot10^{-4} \cdot B^2 / T^3$. In general then, $\Gamma_{5}^{(MHD)} \sim 7 \times 10^{-4} \cdot e^6\cdot B^2$, modulo electric charge logarithmic corrections. Alternatively, from the fluctuation-dissipation relation $\Gamma_{\rm diff} = {1\over 6}\Gamma_{5}T^3$, the effective diffusion rate is expected in MHD as
\begin{eqnarray}\label{eq:GammaDiffMHD}
\Gamma_{\rm diff}^{(MHD)}
\simeq 4.1\cdot 10^{-5}\log({17.6/e^2})\,e^6B^2\,,
\end{eqnarray}
which e.g.~for $e^2 = 1$ is $\Gamma_{\rm diff}^{(MHD)} \simeq 1.1\cdot10^{-4} \cdot B^2$, or again in general $\Gamma_{\rm diff}^{(MHD)} \sim 10^{-4} \cdot e^6B^2$ modulo logarithmic corrections on the electric charge.

Comparing these rates against our numerical fits to $\Gamma_5$ (c.f.~Table~\ref{tab:table1}) or against the fits to $\Gamma_{\rm diff}$ from Ref.~\cite{Figueroa:2017hun} (c.f.~Table~\ref{tab:table2}), we observe that the numerically extracted rates are a factor $\sim 10$ larger than the MHD counterparts. In particular for $e^2 = 1$ and imposing an exact scaling $\propto B^2$ over our data (something that it is very well verified, recall the discussion in
 section~\ref{subsec:RateComparison}), we obtain\footnote{The ratio ${\Gamma_{5}^{(num)}/\Gamma_{5}^{(MHD)}}$
 of Eq.~\eqref{eq:ratios} was obtained with the power law fit $\Gamma_{5}^{(num)}\propto e^{11/2}$.
 If the ratio is obtained instead with the logarithmic fit $\Gamma_5^{(num)}\propto e^6\ln\left (\frac{62}{e^2}\right)$, this
yields (for $e^2 = 1$)
 ${\Gamma_{5}^{(num)}/\Gamma_{5}^{(MHD)}} = 11.3\substack{+1.0 \\ -1.4}$,  which is equally consistent with the other ratios.}
\begin{eqnarray}
{\Gamma_{5}^{(num)}\over\Gamma_{5}^{(MHD)}}\Big|_{e^2 = 1} = 11.2\substack{+0.1 \\ -0.1}\,,~~~~~~{\Gamma_{\rm diff}^{(num)}\over\Gamma_{\rm diff}^{(MHD)}}\Big|_{e^2 = 1} = 10.5\substack{+6.5 \\ -4.0}\,.
\label{eq:ratios}
\end{eqnarray}
The two ratios in Eq.~(\ref{eq:ratios}) are, as expected, consistent with each other, even if these are numbers independently obtained. This is simply due to the fact that the fluctuation-dissipation argument relating $\Gamma_5$ and $\Gamma_{\rm diff}$, is actually well verified in our data (within the errors of the diffusion rate), recall section~\ref{subsec:RateComparison}. In summary, both the chiral decay rate or the Chern-Simons diffusion rate, obtained by completely independent simulations, lead to rates much larger than predicted in the MHD picture. \modif{This is the main quantitative result of our work.\footnote{Note that this result can also be interpreted as having an effective conductivity $10$ times smaller than that coming from kinetic theory.}}

\modif{The equations \ref{eq:ratios} lead us to conclude that formula \eqref{eq:chiral_mag_rate} is not describing correctly the rate of fermion number non-conservation in the presence of magnetic fields in our system. We take this as an evidence of the impact of short-scales fluctuations on the system dynamics. Assuming it this to be correct, it calls for a revision of the cosmological implications of fermion number and chirality non-conservation in finite temperature Abelian gauge theories.}

\modif{In fact, the failure of the MHD to predict correctly the CS diffusion rate in the presence of a magnetic field is not surprising. The MHD is only valid at distances larger than the mean free path of particles in the plasma and thus it is not accounting correctly for the shorter scale electromagnetic fluctuations that may change the CS number and which are automatically included in our simulations. The similar phenomena exists in the diffusion of the CS number in non-Abelian theories, where the relevant distance scale $\sim \frac{1}{\alpha T}$ is much smaller than the scattering length $\sim \frac{1}{\alpha^2 T}$. Though our lattice simulations, being classical in nature,  do not include several effects - hard thermal loops (HTL) and scattering of energetic particles with energy $\sim \mathcal{O}(T)$, we do believe that they are reliable. The HTLs  introduce a Debye-screening of the electric fields, plasmon masses for thermal excitations,  and the Landau damping \cite{Braaten:1989mz,Andersen:1999sf}.
 All these effects have the energy scale $\sim e T$. In our discretised system, the similar effects are generated through lattice artifacts \cite{Bodeker:1995pp}.  With our parametrisation, we have a UV cutoff of order $\sim T$, which will induce corrections mimicking  HTLs \cite{Bodeker:1995pp}. Regarding the effect of collisions on the CS diffusion rate, we expect them to be negligible, as the longest scales which enters our problem is smaller than the mean free path. As detailed in section 2.2 of \cite{Figueroa:2017hun}, the length scale associated to the chemical potential decay is the typical size of a configuration carrying CS-number one $l_{CS}\sim \frac{1}{\alpha T}$,
  while the mean free path in the plasma is of order $l_{mf}\sim \frac{1}{\alpha^2 T}$. These conjectures can be checked by including HTLs and effective collision terms into classical equations of motion along the lines of ref. \cite{Bodeker:1999gx,Rajantie:1999mp,Hindmarsh:2001vp}. We leave this for the future work.}

\section{Conclusions}
\label{sec:disc}

In this work, we studied the evolution of the fermionic charge in an Abelian gauge theory at finite temperature, which can be a proxy either for fermion non-conservation or for chirality breaking.
In section \ref{sec:chem_pot_finite_vol}, we studied the dependence of the evolution on the initial value of the chemical potential. We observed that for small chemical potential, it develops long initial $plateau$s, before the decay is triggered. We saw in section~\ref{sec:inv_cascade} that for larger initial chemical potential, the decay happens through a self-similar process. This leads to a phenomenon of inverse cascade in the gauge field sector, with power transferred from UV to IR scales. We observed and quantified both of these phenomena, measuring the critical exponent of the chemical potential self-similar decay to be $-\frac{1}{2}$.

In section \ref{sec:ov_dyn_B}, we moved on to
the study  of the effect of an external magnetic field on the chemical potential evolution. Both for large and small initial chemical potential, three different situations can happen. First, when the external magnetic field is small, the chemical potential relaxes to its critical value, in the same way that in the absence of chemical potential. Then, from the critical value, it eventually decays to zero. For large external magnetic field, the dynamic is fully determined by the magnetic field, and the decay happens through damped oscillations. For moderate magnetic fields, both effects can be observed at the same time.

A way to isolate the effect of the external magnetic field influence is to study configurations with initial chemical potential equal to the critical one, as presented in \ref{subsec:RateComparison}. Then, only the exponential decay is visible. Following this procedure, we could extract a chiral dissipation rate and study its dependence on the parameters of the theory, finding $\Gamma_5=10^{-2.10\pm0.02}B^2 e^{11/2}/T^3$.
We also compared this rate to the fluctuation rate of the topological charge in the absence of chemical potential, which was measured in Ref.~\cite{Figueroa:2017hun}. These rates are related through a fluctuation-dissipation theorem. We found a good agreement between them, providing a solid self-consistency check for our framework.

In section \ref{sec:compMHD} we analysed our results in light of MHD, which is the long-wavelength effective theory of our system. Despite being out of its range of validity, qualitative features of the chemical potential can be well described. For example, in section \ref{subsec:plateau}, we provide a simplified MHD modeling which reproduces the time-scale of the observed $plateau$s in the chemical potential decay. In section \ref{subsec:MHD_inv_casc} we also showed that the spectrum evolution of the inverse cascade is well-modeled by a MHD-inspired description.
\modif{In section  \ref{subsec:MHD_rates}, we studied the chiral rate in the physically relevant range of chemical potentials. We showed that our numerical rates are an order of magnitude larger than the MHD predictions. We interpreted this as evidence of the effect of short-scale fluctuations on the system dynamics. This can potentially shed a new light on the role of the abelian contribution to the anomaly on fermion number/chirality violating processes.}

\modif{The main outlook of this work will be to confirm or refute these results. We intend to verify numerically the effects of hard thermal loops and collisions, using some effective theory, see e.g.~\cite{Bodeker:1999gx,Rajantie:1999mp,Hindmarsh:2001vp}. }

\modif{Another direction which remains to be explored is the regime of extremely small chemical potential.}
This is however challenging from a technical point of view, as the smaller the initial values of the chemical potential, the larger the volume needs to be, given the existence of a critical value $\mu_c$. Moreover, one also need longer simulations, as the initial $plateau$ gets longer. Our computer resources thus limits us to explore systematically this regime.

\FloatBarrier

\section*{Acknowledgements}

The authors want to thank  Oleksandr Sobol for relevant comments and Kohei Kamada for useful discussions. A.F. wishes to thank Hauke Sandmeyer for giving him access to  his numerical fitting routines and Pavel Buividovich for an insightful discussion.  This work was supported by the ERC-AdG-2015 grant 694896 and the Swiss National Science Foundation. The numerical simulations were performed on the Intel Broadwell based cluster Fidis, provided by the EPFL HPC center (SCITAS). The discrete Fourier transforms were performed with the AccFFT library \cite{2015arXiv150607933G}.

\appendix

\section*{Appendices}
\section{Lattice set-up}
\label{app:latDesc}

The results presented in this work were obtained using the lattice discretisation presented in Ref.~\cite{Figueroa:2017qmv}. Initial conditions for the gauge fields and the scalar fields are drawn from a thermal ensemble, generated by a simple Metropolis algorithm. Gauss law is then enforced on the thermalised configurations. After that, the system is evolved along the classical trajectory specified by the set of discretised equations of motion

\begin{align}
    \pi &= \Delta_o^+\varphi &\Delta_o^-\pi &= \sum_iD_i^-D_i^+\varphi - V_{,\varphi^*}\\
    E_i &= \Delta_o^+A_i &\Delta_o^-  E_i &= {2e},{\rm Im}\lbrace\varphi^*D_i^+\varphi\rbrace-\sum_{j,k}\epsilon_{ijk}\Delta_{j}^-B_k- {e^2\over 4\pi^2}\mu B_i^{(8)} \\
    \mu &= \frac{T}{2\sqrt{3}}\Delta_o^-a & \Delta_o^+\mu &= {3\over \pi^2} {1\over T^2}{e^2\over N^3}\sum_{\vec n} {1\over2}\sum_i E_i^{(2)}(B_i^{(4)}+B_{i,+0}^{(4)}) \\
    \sum_i \Delta_i^- E_i &= {2e}\,{\rm Im}\lbrace\varphi^*\pi\rbrace
\end{align}
with
\begin{align}
   E_i^{(2)} &\equiv {1\over2}(E_i+E_{i,-i}) &  B_i^{(4)} &\equiv {1\over4}(B_i+B_{i,-j}+B_{i,-k}+B_{i,-j-k})\\
  B_i^{(8)} &\equiv {1\over2}\left(B_i^{(4)}+B_{i,+i}^{(4)}\right)\,,
\end{align}
and  $\Delta^{\pm}_\mu f=\pm\frac{1}{\rm dx}(f_{\pm\mu}-f)$, $D_\mu^\pm f =\pm\frac{1}{\rm dx}(e^{\mp i e {\rm dx^\mu}A_\mu(n\pm\frac{1}{2})} f_{\pm\mu}-f)$ the forward/backward finite difference operator and covariant derivatives. The notation $f_{a,\mu}$
means that the component $a$ of the vector field $f$ is to be evaluated at the point $\vec{n}+\hat{\mu}$, with $\hat{\mu}$ a unit displacement in the direction $\mu$, $f_{a,\mu}=f_a(n+\hat{\mu})$. Notice that the equations are built out of composite field so that all the fields can be expanded consistently about the same point to reproduce the continuum equations to order $O(\dd x^2)$. The scalar fields live on lattice edges while the gauge fields are link variables; they live on half-integer sites. The time differential operators evolve by half a step in times. More details are to be found in \cite{Figueroa:2017qmv}.

The constant background magnetic field is introduced through twisted boundary conditions~\cite{Kajantie:1998rz}, which imposes a constant flux. To specify, we modify the periodic boundary conditions of the first component of our gauge field as follow

\begin{equation}
  A_1(1,N,n_3)=A_1(1,0,k)- \frac{2\pi n_{mag}}{\dd x} \ \ \ \forall k\in[0,N-1] .
  \label{eq:twist_bc}
\end{equation}
This corresponds to introducing a constant background magnetic field in the $z$-direction of magnitude $B=\frac{2\pi n_{mag}}{(\dd x N)^2}$.

For large volumes, it is essential to initiate the Monte-Carlo with configurations which already satisfy the twisted-boundary conditions \eqref{eq:twist_bc}. This can be achieved by taking as seed

\begin{align}
  A^{init}_1(i,j,k)&=-\frac{2\pi n_{mag}}{\dd xN^2} j -\delta_{j0}\frac{2\pi n_{mag}}{\dd xN}+ \delta_{j0}\delta_{i1}\frac{2\pi n_{mag}}{\dd x}   \label{eq:twist_init_1}
\\
  A^{init}_2(i,j,k)&=\delta_{j0}\frac{2\pi n_{mag}}{\dd xN} i -\delta_{j0}\frac{2\pi n_{mag}}{\dd x}+\delta_{j0}\delta_{i0}\frac{2\pi n_{mag}}{\dd x}+\delta_{j0}\delta_{i1}\frac{2\pi n_{mag}}{\dd x}  \label{eq:twist_init_2}
\\
  A^{init}_3(i,j,k)&=0,
  \label{eq:twist_init_3}
\end{align}
where the non-trivial dependence of $A_2$ ensures that $A$ is consistent. In more details, a gauge field which lives on a periodic manifold must transform as
\begin{align}
  A_\mu(x+N\cdot \hat{i})=A_\mu(x)+\partial_{\mu}\alpha_i(x)\,,
\end{align}
with potentially three different gauge transformations $\alpha_i$, one by direction. These gauge transformations cannot be completely arbitrary, they satisfy so-called compatibility conditions. These are the requirement that the relation of the gauge field value at a given point $x+N\cdot \hat{i}+N\cdot\hat{j}$ to its value at $x$ better not depend on whether one first goes through the $i$ or through the $j$ boundary. In other words, we have

\begin{align}
  A_{\mu}(x+N\cdot \hat{i}+N\cdot\hat{j})=A_{\mu}(x+N\cdot \hat{j}+N\cdot\hat{i})
  \label{eq:comp_cont}
\end{align}
which implies
\begin{align}
  \partial_\mu(x)\alpha_i(x)+\partial_\mu\alpha_j(x+N\cdot \hat{i}) = \partial_\mu\alpha_j(x)+\partial_\mu\alpha_i(x+N\cdot \hat{j})
\end{align}
The twisted boundary conditions \eqref{eq:twist_bc} tells us that $\partial_1 \alpha_2(1,j,k)=-\frac{2\pi n_{mag}}{\dd x}$. The choice of initial field \eqref{eq:twist_init_1}~-~\eqref{eq:twist_init_3} is made to satisfy the compatibility conditions \eqref{eq:comp_cont}.

\section{Fluctuation-dissipation theorem}
\label{app:fluc-diss}
The dissipative chiral decay rate $\Gamma_5$ is related to the diffusion rate of the topological charge $\langle Q^2 (t)\rangle = \Gamma V t$. To derive this relation, we will use Zubarev's formalism \cite{zubarev1974nonequilibrium}, in the spirit of Ref.~\cite{Khlebnikov:1988sr}. We treat $\mu$ as a dynamical variable and want to understand its out of equilibrium properties. To do so, we introduce a "local equilibrium" partition function

\begin{equation}
  \rho_{LE}=\exp\left ( -\beta H + VT \chi(t) \mu(t)\right )
\end{equation}
with $\chi(t)$ some Lagrange parameter which drives $\mu(t)$ locally out of equilibrium and the factor $VT^2$ was introduced to make $\chi(t)$ of the same dimensions than $\mu$. The function $H$ is the Hamiltonian associated with our system. The relevant $ \mu$-dependent part is

\begin{align}
H_\mu= \frac{1}{24}\mu^2 V T^2
\end{align}

This function is used to compute $\langle \mu \rangle$ at any time \cite{zubarev1974nonequilibrium} and thus determines $\chi(t)$ as a function of $\langle \mu \rangle$.

\begin{align}
  \langle\mu\rangle = \frac{\int_{-\infty}^\infty\dd \mu \ \mu \exp\left ( -\frac{1}{24}\mu^2 V T + \chi\mu  V T \right)}{\int_{-\infty}^\infty\dd \mu \exp\left ( -\frac{1}{24}\mu^2 V T  + \chi \mu   V T \right)}
  =12\chi
  \label{eq:chimu}
\end{align}
However, it is \textit{not} stationnary, i.e. it does not satisfy Liouville's equation

\begin{equation}
  \frac{\dd \rho}{\dd t}=\frac{\partial \rho}{\partial t}+ \{\rho,H\}=0 ,
\end{equation}
with $\langle \cdots\rangle$ the usual Poisson brackets. A way to fix this is to introduce a second density matrix
\begin{equation}
  \rho = N\lim_{\epsilon\to 0^+} \exp\left ( -\frac{H}{T}+V T^2\epsilon \int_{-\infty}^te^{T\epsilon(t-t')}\chi(t')\mu(t')\dd t'\right )
\end{equation}
with $N$ such that $\rho$ integrates to one. Integrating by part and neglecting $\dot{\chi}(t)$, i.e. considering slow processes, this can be recast as
\begin{align}
  \rho &= N\lim_{\epsilon\to 0^+} \exp\left ( -\frac{H}{T}+ VT \chi  \mu - V T  \int_{-\infty}^te^{T\epsilon(t-t')}\chi(t')\dot{\mu}(t')\dd t'\right )\\
  &\equiv N(h)\lim_{\epsilon\to 0^+} \exp\left ( -\frac{H}{T}+h\right )
\end{align}
with $h=VT \chi  \mu - V T  \int_{-\infty}^te^{T\epsilon(t-t')}\chi(t')\dot{\mu}(t')\dd t'$. Working at linear order in $h$, it becomes
\begin{align}
  \rho &=\lim_{\epsilon\to 0^+}\rho_0(1+h+\langle h\rangle_0 )\\
  &\approx \lim_{\epsilon\to 0^+}\rho_0(1+VT \chi  \mu - V T  \int_{-\infty}^te^{T\epsilon(t-t')}\chi(t')\dot{\mu}(t')\dd t'+\langle h\rangle_0 ) \label{eq:part_func_lin_h}
\end{align}
where $\rho_0=N_0 e^{-\frac{H}{T}}$ and $\langle h\rangle_0$ is the average with respect to $\rho_0$.

To obtain a relation between the chiral decay rate and the diffusion rate of the topological charge, we compute $\langle\dot{\mu}\rangle$ using eq. \eqref{eq:part_func_lin_h}. Using the fact that in equilibrium we have  $\langle\dot{\mu}\rangle_0=0$ and $\langle\dot{\mu}\mu\rangle_0=0$, we obtain

\begin{equation}
\langle\dot{\mu}\rangle=-VT \chi(t)\lim_{\epsilon\to 0^+}\left <\int_{-\infty}^te^{T\epsilon(t-t')}\dot{\mu}(t')\dot{\mu}(t)\dd t'\right >_0
\end{equation}
where again we assumed that $\chi(t)$ as a weak dependence on time. Now we can make use of the anomaly equation \eqref{mu5eq}

\begin{align}
\langle\dot{\mu}\rangle&=-VT \chi(t)\left(\frac{3e^2}{4T^2\pi^2} {1\over V}\right )^2\cdot\notag\\
&\lim_{\epsilon\to 0^+}\left <\int_{-\infty}^te^{T\epsilon(t-t')}\int d^3x\,F_\mn \tilde{F}^\mn(x,t')\int d^3xF_\mn \tilde{F}^\mn(y,t)\dd t'\right >_0\\
&=- \chi(t) \frac{9\cdot 16}{T^3 V}\lim_{\epsilon\to 0^+}\left <\int_{-\infty}^te^{T\epsilon(t-t')}q(t')q(t)\dd t'\right >_0,
\end{align}
with $q(t)=\frac{e^2}{16\pi^2}\int d^3x\,F_\mn \tilde{F}^\mn(x,t)$ the topological charge density. With this notation and setting taking $t=0$ as a reference time, we can define the topological charge as

\begin{equation}
  Q(t-t_0)=\int_{t_0}^{t} \dd t'q(t') .
\end{equation}
Inserting Eq. \eqref{eq:chimu} and setting $\epsilon=0$, we then have

\begin{align}
\langle\dot{\mu}\rangle
&=- \langle \mu\rangle \frac{12}{T^3 V}\lim_{t''\to \infty}\left < Q(t'')q(0)\right >_0,
\label{eq:kinetic}
\end{align}
This last quantity also appears in the diffusion rate of the topological charge. In the absence of chemical potential, $Q(t)$ follows a random walk and we can define its diffusion rate by, following the convention of \cite{Figueroa:2017hun}

\begin{equation}
  \langle Q^2(t-t_0)\rangle =\Gamma_{\rm diff} V (t-t_0),
\end{equation}
for times $t$ much larger than a reference time $t_0$\footnote{In equation \eqref{eq:diff_rate}, we implicitly set $t_0=0$.}.
We can differentiate this expression by $t_0$ to get

\begin{align}
-2\langle  Q(t-t_0) q(t_0)\rangle &=-\Gamma_{\rm diff} V \\
\implies \langle Q(t-t_0)q(t_0)\rangle &=\frac{\Gamma_{\rm diff}V}{2}
\end{align}
Finally, plugging this in \eqref{eq:kinetic}, we get

\begin{align}
\langle\dot{\mu}\rangle
&=- \langle \mu\rangle \frac{6}{T^3 V}\Gamma_{\rm diff}V,
\label{eq:kinetic}
\end{align}
Leading to the relation
\begin{equation}
  \Gamma_5=6\frac{\Gamma_{\rm diff}}{T^3}.
\end{equation}




\providecommand{\href}[2]{#2}\begingroup\raggedright\endgroup
\end{document}